\documentclass[sigconf,letterpaper,natbib=false]{acmart}

\usepackage{makecell}
\usepackage[linesnumbered,boxed]{algorithm2e}
\usepackage{subcaption}
\usepackage{xspace}
\usepackage{multirow}
\usepackage{pifont}
\usepackage{colortbl}

\def\Snospace~{\S{}}
\hypersetup{
    colorlinks,%
    citecolor=[rgb]{0.7,0,0},
    filecolor=[rgb]{0.2,0.2,0.2},
    linkcolor=[rgb]{0,0,0.7},
    urlcolor=[rgb]{0.7,0,0},
}

  \newcommand{\USC}{USC\xspace}

\makeatletter
  \newcommand\EatSpacesHack{\@bsphack\@esphack}
\makeatother

  \renewcommand\comment[1]{\EatSpacesHack}
  \newcommand\PostSubmission[1]{\EatSpacesHack}
  \newcommand\STV[1]{\EatSpacesHack}
  \newcommand\JH[1]{\EatSpacesHack}

  \newcommand\reviewfix[1]{\EatSpacesHack}

\newcommand{\ResRouter}[1]{Residence #1\xspace}

\makeatletter
  \renewcommand\section{\def\@toclevel{1}%
    \@startsection{section}{1}{\z@}%
    {-.4\baselineskip \@plus -2\p@ \@minus -.2\p@}%
    {.2\baselineskip}%
    {\ACM@NRadjust\@secfont}}
  \let\ACM@origsection\section
  \renewcommand\subsection{\def\@toclevel{2}%
    \@startsection{subsection}{2}{\z@}%
    {-.25\baselineskip \@plus -2\p@ \@minus -.2\p@}%
    {.1\baselineskip}%
    {\ACM@NRadjust\@subsecfont}}
  \let\ACM@origsubsection\subsection
\makeatother

  \copyrightyear{2025}
  \acmYear{2025}
  \setcopyright{cc}
  \setcctype{by-nc-sa}
  \acmConference[IMC '25]{Proceedings of the 2025 ACM Internet Measurement Conference}{October 28--31, 2025}{Madison, WI, USA}
  \acmBooktitle{Proceedings of the 2025 ACM Internet Measurement Conference (IMC '25), October 28--31, 2025, Madison, WI, USA}
  \acmDOI{10.1145/3730567.3764467}
  \acmISBN{979-8-4007-1860-1/2025/10}

\RequirePackage[datamodel=acmdatamodel,style=acmnumeric]{biblatex}

\begin{document}

\title{Towards a Non-Binary View of IPv6 Adoption}

  \author{Sulyab Thottungal Valapu}
  \orcid{0000-0001-6332-1344}
  \affiliation{
      \institution{University of Southern California}
      \department{Thomas Lord Dept.~of Computer Science}
      \city{Los Angeles}
      \country{United States}
  }
  \email{sulyab.tv@usc.edu}
  
  \author{John Heidemann}
  \orcid{0000-0002-1225-7562}
  \affiliation{
      \institution{USC/Information Sciences Institute}
      \department{Thomas Lord Dept.~of Computer Science}
      \city{Los Angeles}
      \country{United States}
  }
  \email{johnh@isi.edu}

%
%
%
%
%
%
\renewcommand{\shortauthors}{Sulyab Thottungal Valapu and John Heidemann}

%
%
%
\begin{abstract}
Twelve years have passed since World IPv6 Launch Day,
  but what is the current state of IPv6 deployment?
Prior work has examined IPv6 status as a binary:
  can a user do \emph{any} IPv6?
As deployment increases,
  we must consider a
  more nuanced, non-binary perspective on IPv6:
  \emph{how much and often can a user or a service use IPv6?}
We consider this question as a client, server, and cloud provider.
Considering the client's perspective,
  we observe user traffic.
We see that the fraction of IPv6 traffic a user sends varies 
  greatly, both across users and day-by-day,
  with a standard deviation of over 15\%.
We show this variation occurs for two main reasons.
First, IPv6 traffic is primarily human-generated, thus showing diurnal patterns.
Second, some services lead with full IPv6 adoption, while others lag with
  partial or no support,
  so as users do different things their fraction of IPv6 varies.
We look at server-side IPv6 adoption in two ways.
First, we expand analysis of web services to examine how many are only partially IPv6 enabled due to their reliance on IPv4-only resources.
Our findings reveal that only 12.6\% of top 100k websites %
   qualify as fully IPv6-ready.
Finally, we examine cloud support for IPv6.
Although all clouds and CDNs support IPv6, we find that tenant deployment rates vary significantly across providers.
We find that ease of enabling IPv6 in the cloud is correlated with tenant IPv6 adoption rates, and recommend best practices for cloud providers to improve IPv6 adoption.
Our results suggest IPv6 deployment is growing, 
  but many services lag, presenting a potential for improvement.
\end{abstract}


%
%
%

%
%
%

%
%
%
\maketitle

\section{Introduction}
	\label{sec:intro}

More than a decade has passed since World IPv6 Launch Day in 2012~\cite{ISOC12a}.
IPv6 is prominent in many mobile operators,
  and recently Vietnam announced all of their networks
  will support IPv6~\cite{dobberstein_vietnam_2024}.
IPv6 is clearly past the ``early adopter'' phase
  of the technology lifecycle~\cite{Beal57a},
  but it also is not ubiquitous.
How do we judge where IPv6 is today?

Prior studies have evaluated IPv6 maturity in several ways.
Early evaluation asked the \emph{binary} question: \emph{can one access IPv6}?
For example,
  first evaluations as IPv6 moved from design to deployment
  asked which countries supported IPv6~\cite{white_internet_2005}.
More recent work looks inside countries
  to explore what ISPs or Autonomous Systems offer IPv6~\cite{InternetSocietyPulse}.
Alternatively, one can summarize this evaluation
  as the fraction of users in a country with access,
  based on ISP market share.
For example, Internet Society reports that 73\% of users from India have IPv6 access~\cite{InternetSocietyPulse}.
These questions are \emph{necessary}---an individual cannot access IPv6 if their country or ISP does not support it.
But they are not \emph{sufficient}.
Your country or ISP being ``IPv6 ready'' is quite different from \emph{actual} IPv6 prefix delegation to specific users in all regions, and that is different from \emph{use} of IPv6 by those users.

\textbf{Contributions:}
The first contribution of this paper
  is to \emph{re-frame IPv6 deployment in non-binary terms}.
Rather than simply asking, ``is IPv6 possible?'',
  we measure how much traffic is actually IPv6?
Following Metcalfe's Law~\cite{metcalfe_metcalfes_2013}, 
  value in networking is not a function of \emph{me},
  but a function of \emph{us}---me and those with whom I interact.
Our
  non-binary view of IPv6 adoption
  leads us to the more nuanced question of
  \emph{how often do users and services interact using IPv6},
  beyond the prior: is it possible for a user or service to use IPv6 at all.
This re-framing is of growing importance
  as IPv6 support grows,
  since the binary ``is it possible'' view no longer gives new insight
  since the answer is almost always ``yes''.
A practical benefit of understanding IPv6 transition today
  is to help evaluate the importance
  today's markets built around IPv4 scarcity,
  such IPv4 marketplaces~\cite{livadariu_first_2013,prehn_when_2020}
  and Amazon's 2023 shift to charging for IPv4 addresses~\cite{barr_new_2023,khan_amazon_2024}.

To develop this non-binary view, our second contribution
  is to design measurements to  consider ``how much IPv6'' from three different perspectives.
For \emph{users}, how much of their traffic is actually IPv6?
Which services that they regularly use lag with only IPv4 support,
  holding back IPv6 from 100\%?
For \emph{services}, how complete is a website's support of IPv6?
How many of its embedded resources are available over IPv6, and which widely used resources remain IPv4-only?
For \emph{cloud providers}, what fraction of tenants support IPv6?
While all clouds offer IPv6 support, do they differ in how easily it can be enabled?

Our final contribution is the results of these measurements.
We study the traffic of five residences,
  finding large variation in IPv6 traffic
  by residence and by day, driven by human activity and showing a clear diurnal pattern.
Our investigation of the top 100k websites
  shows that, of the 34,661 sites that are reachable via IPv6,
  nearly three-fourths of these sites depend on resources only accessible by IPv4,
  leaving only 10,308 (12.6\%) websites fully functional over IPv6.
Our analysis of cloud platforms finds substantial differences in tenant IPv6 adoption, closely tied to how easily IPv6 can be enabled on each provider.
We conclude with recommendations for cloud providers to increase IPv6 adoption by improving defaults and minimizing configuration burden.

\textbf{Ethical Considerations and Data Availability:}
Our user studies in this paper were IRB-reviewed 
  (USC IRB \#UP-24-00738)
and done with user consent; we expand on ethical considerations in \autoref{sec:appendix/ethics}.
Our client-side data is not available because of anonymization requirements.
Our server-side and cloud data are available online at \url{https://ant.isi.edu/datasets/ipv6}.

\section{Related Work}
    \label{sec:related}

Efforts to measure the progress %
  of IPv6 adoption have %
  been ongoing for more than a decade.
IPv6 adoption has been examined along multiple dimensions, including end-user readiness, service availability, and organizational deployment.

\textbf{IPv6 Use by Users:}
One major thread of research measures the extent to which end users have IPv6 connectivity.
Early work by Karpilovsky et al.~found that IPv6 traffic was primarily limited to control traffic (DNS and ICMP)~\cite{karpilovsky_quantifying_2009},
  but later studies reported increasing volumes of user-generated traffic~\cite{sarrar_investigating_2012,li_towards_2020}.
Pujol et al.~observe that actual IPv6 usage remained low in a dual-stack ISP, often due to IPv6-incompatible customer premises equipment (CPE)~\cite{pujol_understanding_2017}.
Colitti et al.~\cite{colitti_evaluating_2010} and later Zander et al.~\cite{zander_mitigating_2012} measure end-user IPv6 capability by embedding JavaScript fragments in web pages, with the latter study reporting greater IPv6 availability at workplaces than at residences.
However, Li's later study finds the opposite~\cite{li_towards_2020}.
Google's IPv6 adoption tracker shows that the share of users that access Google over IPv6 grew from under 1\% in 2012 (World IPv6 Launch) to nearly 50\% by 2025~\cite{Google22a}.
However, statistics from AMS-IX, one of the world's largest Internet exchanges, show that IPv6 still accounts for only about 8\% of total traffic in 2025~\cite{AMS_IX}.

\emph{Our contribution:}
We introduce a non-binary perspective on IPv6 use by end-users in~\autoref{sec:client} to explain the differences between IPv6 \emph{capability},
  as measured by Google's user-to-Google statistics,
  and \emph{aggregate usage}, as reflected in the many-to-many traffic fraction at AMS-IX.
We look at real user traffic from multiple residences,
  focusing on which specific applications drive IPv6 and which generate traffic that remains IPv4-only.

\textbf{IPv6 Use by Services:}
Another line of work examines how widely online services have deployed IPv6.
Nikkhah et al.~evaluate IPv6 accessibility and performance for Alexa's Top 1M websites, attributing IPv6's inferior performance to control-plane differences~\cite{nikkhah_assessing_2011}.
Dhamdhere et al.~confirm that data-plane performance was comparable when AS paths matched~\cite{dhamdhere_measuring_2012}.
Czyz et al.~find that only 3.2\% of Alexa Top 10k sites were reachable via IPv6 in 2014~\cite{czyz_measuring_2014}.
However, these studies focus either on the main page~\cite{nikkhah_assessing_2011,czyz_measuring_2014} or limit themselves to first-party resources~\cite{dhamdhere_measuring_2012}, 
  without evaluating whether the multiple pages on the websites are \emph{fully} IPv6-enabled.
Closest to our work, Bajpai and Sch\"onw\"alder examine the Alexa Top 100 websites and resources embedded \emph{directly} in the main page (including third-party),
finding that 27\% of IPv6-enabled websites included one or more IPv4-only resources~\cite{bajpai_longitudinal_2019}.

\emph{Our contribution:}
We extend Bajpai and Sch\"onw\"alder's methodology two ways.
We use full, browser-based page loads
  that resolve dependencies to arbitrary depth,
  rather than limiting analysis to directly embedded resources (\autoref{sec:serverside}).
We also simulate user interaction by clicking on links to destinations within the same domain, capturing a broader and more realistic set of third-party dependencies.
Our dataset covers 100k websites,
  $1000\times$ more than Bajpai et al.,
  and we compare results over two different years. %
We also perform a more detailed analysis of the types of resources that remain IPv4-only.

\textbf{IPv6 Use by Organizations:}
A third dimension of IPv6 adoption research focuses on deployment by network infrastructure and organizations.
Karpilovsky et al.~showed that although IPv6 prefix allocation was growing exponentially, nearly half of the allocated prefixes were never announced~\cite{karpilovsky_quantifying_2009}.
Dhamdhere et al.~found that while most core Internet transit providers had adopted IPv6, edge networks lagged behind~\cite{dhamdhere_measuring_2012}---a finding reaffirmed five years later~\cite{jia_tracking_2019}.
Streibelt et al.~examined whether DNS infrastructure functions in IPv6-only environments and found that 10\% of DNS providers account for over 97.5\% of all zones that fail to resolve over IPv6~\cite{streibelt_how_2023}.

\emph{Our contribution:}
Given the dominant role of CDNs and cloud platforms in today's Internet---accounting for over half of all web traffic~\cite{sumits_internet_2017} and used by 90\% of companies~\cite{krivec_cloud_2024}---we analyze, in~\autoref{sec:cloud}, how major cloud providers and their tenants are faring in terms of IPv6 adoption.
We also explore how cloud platforms can better support IPv6 deployment.
To our knowledge, we are the first to examine IPv6 adoption in cloud context.

Our most important difference is to pose the new perspective of looking
  at non-binary IPv6 use, rather than 
  basic availability or if a given query can be done completely over IPv6.
Clearly IPv4 will be with us for many, many years to come,
  so our view is that today's focus should be on 
  identifying lagging
  IPv4-only services,
  so we can understand how they can add IPv6 support,
  or when they will sunset.

\section{Client-side Adoption of IPv6}
	\label{sec:client}

Our first goal is to understand
  how much \emph{actual} IPv6 a dual-stack client sends and receives,
  and what factors affect that fraction.

Towards this goal, we observe all Internet traffic 
  from routers at five residential locations 
  in the Los Angeles area,
  \ResRouter{A} through \ResRouter{E}.
A total of 17 individuals live in these residences,
  with one to seven people per household (details omitted due to IRB).
All residences except \ResRouter{B} use Spectrum as their ISP,
  which provides native IPv6 connectivity.
\ResRouter{B} uses Frontier, an IPv4-only ISP;
  IPv6 connectivity there is provided via a tunnel to our university.
Both ISPs are large U.S.~broadband providers.

Each residence has multiple wired and wireless client devices.
At Residences A and B, we manually verify
  that major devices (PCs, phones, tablets, TVs, consoles) 
  are dual-stack, but do not perform this check at other locations.
We recommend residents use the network as they normally would; 
  without any restrictions or specific guidance.

We monitor all inbound and outbound traffic
  for nine months, November 2024 through August 2025.
We observe about 21.5\,TB of traffic over this period;
  \autoref{tab:client-overall} shows traffic volumes and other
  basic statistics by residence.

\begin{table*}
  \centering

  \resizebox{0.85\textwidth}{!}{
    \begin{tabular}{c l 
                    r r r 
                    r r@{\hspace{5pt}}l 
                    r r r 
                    r r@{\hspace{5pt}}l}
      \toprule

      \multicolumn{2}{c}{\textbf{Residence}} 
      & \multicolumn{3}{c}{\textbf{Traffic Volume (GB)}} 
      & \multicolumn{3}{c}{\textbf{Fraction IPv6 (Bytes)}} 
      & \multicolumn{3}{c}{\textbf{Flow Count (M)}} 
      & \multicolumn{3}{c}{\textbf{Fraction IPv6 (Flows)}} \\
      
      \multicolumn{2}{c}{\textbf{and Scope}} 
      & \textbf{Total} & \textbf{IPv4} & \textbf{IPv6} 
      & \textbf{Overall} & \multicolumn{2}{c}{\textbf{Daily mean (s.d.)}} 
      & \textbf{Total} & \textbf{IPv4} & \textbf{IPv6} 
      & \textbf{Overall} & \multicolumn{2}{c}{\textbf{Daily mean (s.d.)}} \\
      
      \midrule

      \cellcolor{white}A & External 
        & 6976.68 & 2242.63 & \textbf{4734.05} 
        & \textbf{0.679} & \textbf{0.686} & (0.173) 
        & 110.61 & 54.95 & \textbf{55.65} 
        & \textbf{0.503} & 0.345 & (0.123) \\
      
      \rowcolor{gray!15}
      \cellcolor{white}  & Internal 
        & 8.87 & \textbf{6.96} & 1.92 
        & 0.216 & 0.225 & (0.114) 
        & 16.38 & \textbf{11.18} & 5.20 
        & 0.318 & 0.318 & (0.121) \\
      
      \midrule

      \cellcolor{white}B & External 
        & 6066.87 & 2197.74 & \textbf{3869.13} 
        & \textbf{0.638} & \textbf{0.549} & (0.202) 
        & 100.65 & 36.96 & \textbf{63.70} 
        & \textbf{0.633} & \textbf{0.650} & (0.110) \\
      
      \rowcolor{gray!15}
      \cellcolor{white}  & Internal 
        & 5.28 & 2.20 & \textbf{3.07} 
        & \textbf{0.583} & \textbf{0.578} & (0.158) 
        & 15.44 & 7.12 & \textbf{8.32} 
        & \textbf{0.539} & \textbf{0.533} & (0.140) \\
      
      \midrule

      \cellcolor{white}C & External 
        & 7816.41 & \textbf{6864.08} & 952.32 
        & 0.122 & 0.089 & (0.188) 
        & 31.71 & \textbf{28.88} & 2.83 
        & 0.089 & 0.116 & (0.106) \\
      
      \rowcolor{gray!15}
      \cellcolor{white}  & Internal 
        & 4.22 & \textbf{2.14} & 2.08 
        & 0.493 & 0.450 & (0.112) 
        & 14.17 & \textbf{9.64} & 4.53 
        & 0.320 & 0.330 & (0.126) \\
      
      \midrule

      \cellcolor{white}D & External 
        & 81.47 & \textbf{41.13} & 40.34 
        & 0.495 & \textbf{0.694} & (0.321) 
        & 1.67 & 0.29 & \textbf{1.37} 
        & \textbf{0.824} & \textbf{0.775} & (0.240) \\
      
      \rowcolor{gray!15}
      \cellcolor{white}  & Internal 
        & 7.18 & 0.10 & \textbf{7.07} 
        & \textbf{0.986} & -- & -- 
        & 10.02 & 0.21 & \textbf{9.81} 
        & \textbf{0.979} & -- & -- \\
      
      \midrule

      \cellcolor{white}E & External 
        & 545.68 & \textbf{509.85} & 35.83 
        & 0.066 & 0.459 & (0.423) 
        & 2.36 & \textbf{2.10} & 0.26 
        & 0.110 & \textbf{0.590} & (0.393) \\
      
      \rowcolor{gray!15}
      \cellcolor{white}  & Internal 
        & 0.26 & \textbf{0.21} & 0.04 
        & 0.173 & 0.396 & 0.369 
        & 0.94 & \textbf{0.77} & 0.18 
        & 0.188 & 0.459 & (0.358) \\
      
      \bottomrule
    \end{tabular}
  }

  \caption{
    Internal and external, per-residence IPv6 traffic volume, flow count, and fraction IPv6 (vs.~IPv4).
  }
  \label{tab:client-overall}
\end{table*}

\subsection{Design of Residence Measurement}
    \label{sec:client/design}

Each residence collects data at its primary router
  to a single upstream ISP\@.
We deploy commodity routers (Linksys E8450) 
  supporting Wi-Fi 6 and wired connections inside the residence.
When possible, we replace the ISP's router,
  but in some cases we place our router downstream (closer to the user)
  of the ISP's router, with devices connecting to our router only.
At \ResRouter{D} and \ResRouter{E}, privacy choices prompt some individuals to prefer to use the ISP router, so we see only part of the household traffic.

On the routers, we deploy OpenWRT (version 22.03.3, running Linux kernel 5.10) extended with a custom built,
  lightweight flow monitor.
Our monitor
  records flow beginnings and ends
  using connection tracking events (\texttt{conntrack}, \texttt{NEW} and \texttt{DESTROY} events)
  from the Linux kernel.
Connection accounting also provides
  information about the flow size in each direction
  (using \texttt{nf\_conntrack\_acct}) %
  with each \texttt{DESTROY} event.
We identify each flow by its 5-tuple of protocol (TCP, UDP, or ICMP)
  and source and destination IP addresses and ports.
For ICMP flows, we record ICMP type, code, and ID.
We log this information at each router
  and send data
  to our server each day.

\subsection{What Fraction of Everyday Traffic is IPv6?}
	\label{sec:client/v6fraction}

\begin{figure}
  \centering

  \begin{subfigure}[t]{0.75\linewidth}
    \centering
    \includegraphics[width=\linewidth]{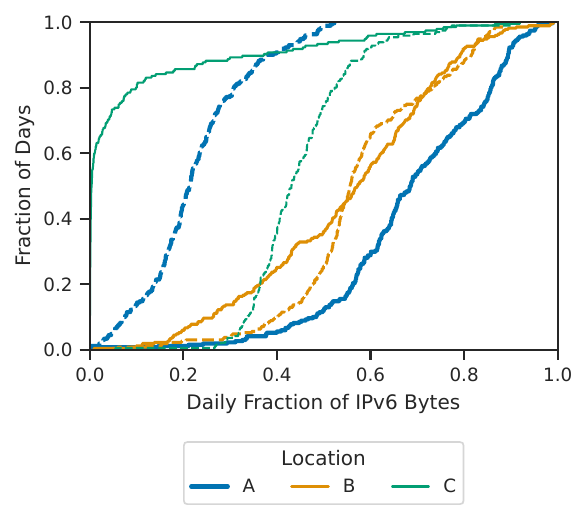}
    \caption{Bytes}
    \label{fig:cdf-v4v6split-bytes}
  \end{subfigure}

  \begin{subfigure}[t]{0.75\linewidth}
    \centering
    \includegraphics[width=\linewidth, trim=0 1.7cm 0 0, clip]{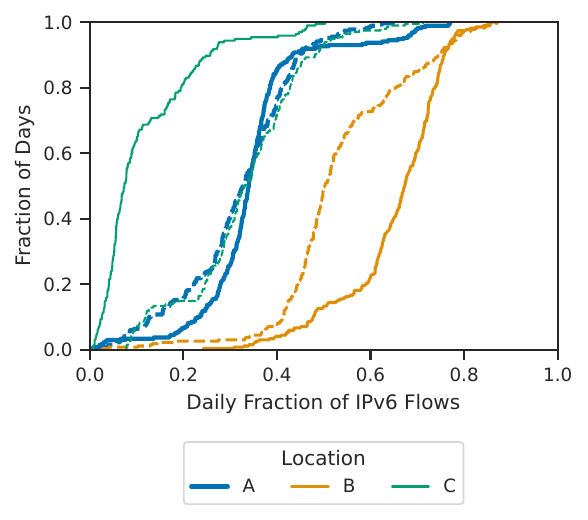}
    \caption{Flows}
    \label{fig:cdf-v4v6split-flows}
  \end{subfigure}

  \caption{
    Fraction of per-day IPv6 bytes and flows (vs. IPv4) at three residences, by external (solid lines) and internal (dashed lines).
  }

  \label{fig:cdf-v4v6split}
\end{figure}

We begin by examining the question: given dual-stack Internet access,
  what fraction of a user's \emph{actual} traffic is IPv6?
We evaluate usage with both counts of unique flows and bytes sent.
Bytes-based metrics emphasize large data exchanges (downloads and streaming media),
  while flow counts emphasize small exchanges such as web surfing.
We report byte-based counts, and add discussion of flow counts where
  they differ.

\autoref{tab:client-overall} shows our observations from 
  our five residences over the nine-month observation period.
\autoref{fig:cdf-v4v6split} shows the daily fraction of IPv6 traffic,
  by both byte volume and flow count, 
  for Residences A, B, and C (the three with the most traffic);
  Residences D and E are in
  \autoref{sec:appendix/other_residences}.

\textbf{Overall, external and internal:}
We first compare overall traffic,
  both external (LAN-to-WAN)
  and internal (LAN-to-LAN) traffic separately.
While we expect most residential traffic to be external, 
  devices in the home may communicate when users play games
  or use network-attached storage, scanners, or printers.

For external traffic, we expect most dual-stack hosts to prefer IPv6 over IPv4,
  since all major web browsers and some other applications
  run the Happy Eyeballs algorithm~\cite{schinazi_happy_2017},
  which attempts IPv6 connections first and falls back to IPv4 only after a brief delay (if IPv6 fails).
Since we verify that major devices at Residences A and B are dual-stack,
  we expect any observed IPv4 traffic at these locations to be for 
  services that are IPv4-only.

Overall, we see that \emph{the majority of flows are IPv6 for most residences} (3 of our 5), whereas \emph{the majority of bytes remain IPv4 for most } (3 of 5).
Our users are shifting to IPv6, but large-volume flows 
  (downloads and streaming) often remain IPv4.
However, fractions vary considerably by residence, as we explore below.

Second, we see that \emph{IPv6 traffic is less used inside the home}, both in terms of flow count and total bytes (2 of 5 residences).
Most residences have little internal traffic (internal is only 1\% of external for 4 of the 5)
  because most devices interact with external servers and not each other.

\textbf{Traffic varies by site:}
We see that \emph{the fraction of external IPv6 varies widely across residences},
  ranging from 0.07 to 0.68 by byte volume and from 0.09 to 0.82 by flow count.
Since Happy Eyeballs prefers IPv6,
  the high level of IPv4 at Residences A and B suggests that many services remain IPv4 only.

Internal traffic shows similarly wide variation
  (0.17 to 0.99 by byte volume and 0.19 to 0.98 by flow count),
  but internal and external IPv6 usage are not well correlated---for instance, Residences C and D have higher internal IPv6 share compared to external,
  whereas Residences A and B show the opposite.

In addition, we see that
  the daily mean fraction of traffic---both by bytes and flow counts---differs
  from the overall mean, suggesting the quantity of traffic varies by day.
This daily variation is also seen by the high standard deviations of daily fractions.

\textbf{Traffic varies by day:}
To understand the high standard deviations
  for day-by-day fractions of IPv6,
  \autoref{fig:cdf-v4v6split} shows the distribution
  of fractions at three residences by external (solid) and internal (dashed).

This result confirms the wide variation in IPv6 traffic over many different days,
  and highlights that even in an IPv6 dominant residence, like
  \ResRouter{A} and \ResRouter{B},
  there are some IPv4-heavy days.

The fraction of IPv6 \emph{bytes} (\autoref{fig:cdf-v4v6split-bytes}) at most residences varies linearly over most of the region,
  with a few heavy-hitter days at the IPv4 or IPv6 end.
We investigated such days, 
  finding that they are often associated with large downloads or video streaming.
Across multiple residences, we frequently observe Valve (AS 32590), Netflix (AS 2906), and Apple (AS 6185) contributing most of the traffic on days with IPv6 fractions above the 90th percentile.
On the other hand, Twitch (AS 46489) and Zoom (AS 30103) dominate traffic on days below the 10th percentile, though these patterns do not appear uniformly across all such days.

In contrast, the fraction of IPv6 \emph{flows} (\autoref{fig:cdf-v4v6split-flows})
  shows low variability, with CDFs that rise sharply over a narrow range,
  indicating that the IPv6 flow fraction tends to remain relatively stable day to day.
We conjecture that this may be due to two factors.
First, some implementations of Happy Eyeballs
  attempt both IPv4 and IPv6 for some or all connections before settling on one.
This behavior suggest that byte-level fractions provide a clearer signal of IPv6 adoption:
  Happy Eyeballs may result in both IPv4 and IPv6 flows being recorded,
  even when nearly all bytes are sent over just one.
Second, the mix of applications used by residents
  tends to remain relatively stable day to day,
  even when \emph{how much use and traffic} from each application varies.
For example, days with heavy video streaming
  show large shifts in traffic volume reflecting video's use or non-use of IPv6,
  while the flow-level statistics may remain stable,
  since the set of applications used remains similar and
  video streams generate large byte volumes per flow.

Overall, this large variation
  by household and day
  suggests that IPv6 use depends 
  heavily on specific services.
High-traffic activities, such as large downloads,
  skew the traffic mix towards or against IPv6 depending
  on that service's protocol support.
This observation prompts us to study 
  IPv6 adoption by services in \autoref{sec:client/leaders-laggards}.

\subsection{Is IPv6 Traffic Periodic?}
   \label{sec:client/temporal}

\begin{figure*}
  \centering
  \includegraphics[width=\linewidth]{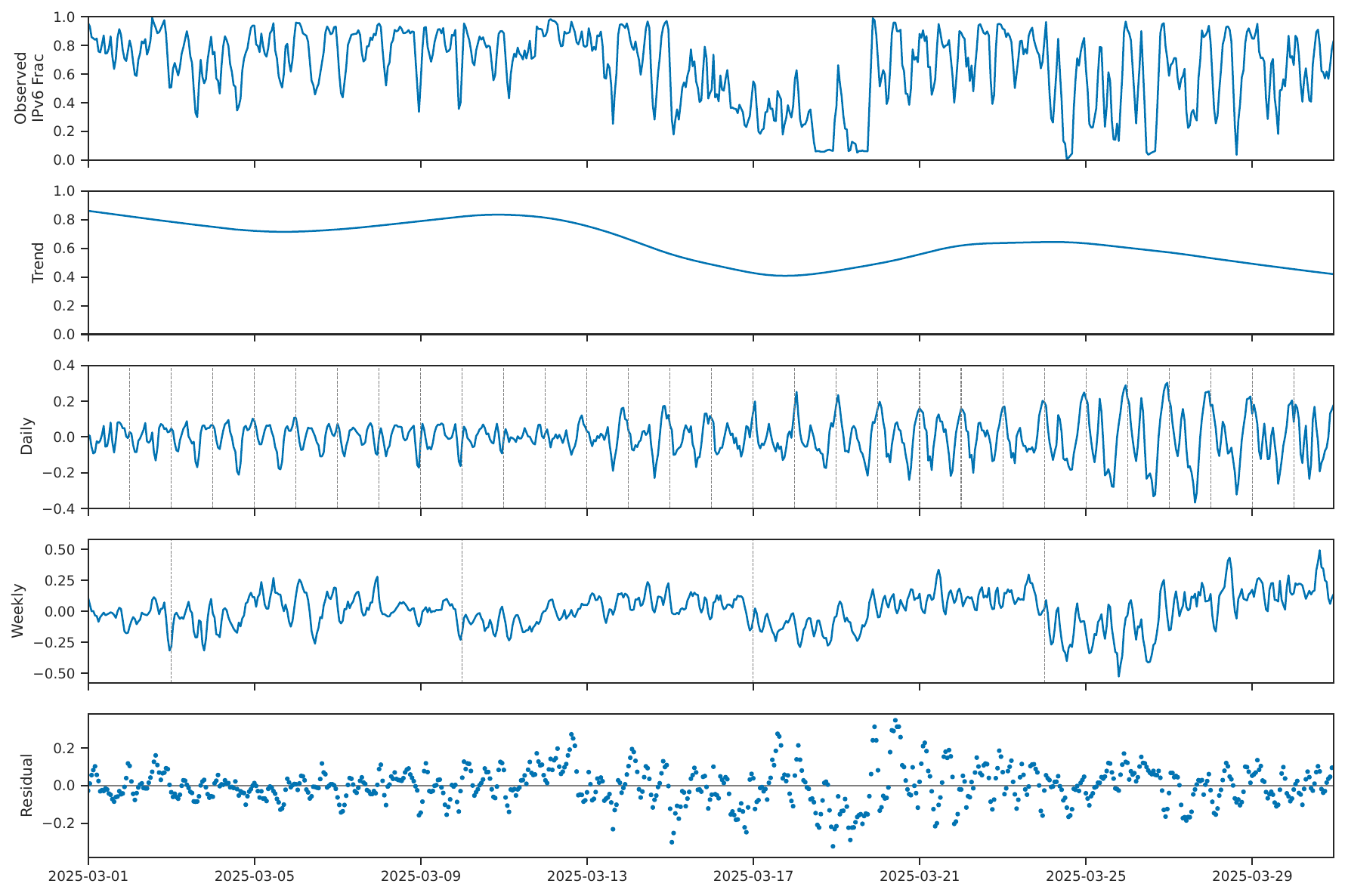}
  \caption{
      Fraction of IPv6 bytes at residence A (top graph), with MSTL decomposition into trend, daily, weekly, and residual components.
  }
  \label{fig:temporal-trends}
\end{figure*}

Most Internet behavior has daily and weekly trends,
  including address use~\cite{Quan14c},
  traffic volume in general~\cite{Papagiannaki05a},
  and IPv6 traffic volume specifically~\cite{Strowes16a}.
Often individuals spend weekdays at work
  and are at home on weekends,
  and user activity follows, with more leisure activities like
  gaming or media consumption on nights and weekends.
We next investigate diurnal and weekly trends
  in our client-side traffic.

Following Baltra et al.\cite{baltra_ebb_2024},
  we apply Multi-Seasonal Trend Decomposition using LOESS (MSTL)~\cite{bandara_mstl_2021} to the IPv6 traffic fraction.
MSTL separates out the overall trend (the long-term running average)
  from daily and weekly trends,
  leaving a residual of variation.

\autoref{fig:temporal-trends} shows the MSTL decomposition
  of the IPv6 byte fraction for external traffic at \ResRouter{A} over March 2025.
Here we show a single month so daily and weekly trends are apparent,
  but we see similar patterns in other months.
We see similar behavior at \ResRouter{B} and \ResRouter{C};
  results are in~\autoref{sec:appendix/mstl-results} for space.
Flow count trends follow the same structure at all three residences;
  the flow-based counterpart to \autoref{fig:temporal-trends} is also available in the appendix.

The top graph shows the daily fraction of IPv6 bytes.
While IPv6 usage is generally high, there is a noticeable drop between March 16 and 19.
This period coincides with the university's spring break,
  when residence A was unoccupied.
This reduction in traffic suggests that IPv6 traffic is primarily human-generated, whereas background traffic tends to be IPv4.

The second graph from the top shows long-term trends.
Again, overall fractions range from 0.4 to 0.8,
  with no long-term direction,
  although a reduction  and recovery of IPv6 mid-March.
We believe this trend varies because the
   usage patterns of individuals vary over time.

The third graph shows the daily component,
  with midnight shown as vertical lines.
We see strong, recurring IPv6 traffic peaks in evenings rising until midnight,
  with a secondary peak mid-morning.
These observations support our observation that IPv6 traffic is largely human-driven,
  reflecting that residents are typically away during the day and return home in the evening.

The fourth panel shows the weekly component,
  with vertical lines on the midnight between Sunday and Monday.
We do not see a strong weekly pattern,
  suggesting that residents are away from home both on weekdays and weekends.

These results provide a client-side view
  that complements prior work showing strong diurnal trends in server-side IPv6 traffic~\cite{Strowes16a}.
While we also observe a diurnal pattern, the weekly pattern is weak.
We conjecture that server-side measurements show greater variation because many users have IPv6 at home but not at work, leading to time-dependent usage.
In contrast, our client-side data---collected from consistently dual-stack environments---suggests that \emph{when IPv6 is always available, usage depends on whether or not the user is at home and active, rather than the time of day or day of the week.}

\subsection{Which Services Lead and Lag in IPv6?}
	\label{sec:client/leaders-laggards}

\begin{figure}
  \centering
  \includegraphics[width=0.8\linewidth]{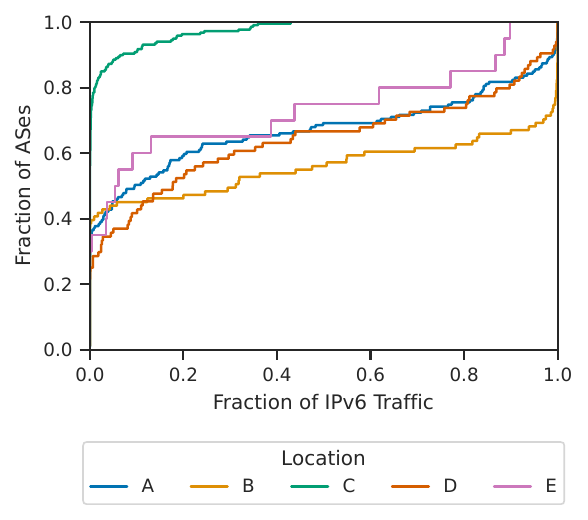}
  \caption{
    Cumulative distribution of IPv6 byte fractions across 35 ASes observed at three or more residences.
    Each point $(x,y)$ on a curve indicates that the IPv6 byte fractions of $y$ fraction of ASes at that location was at or below $x$.
  }
  \label{fig:as-cdf}
\end{figure}

\begin{figure*}
  \centering

  \begin{minipage}[t]{0.48\linewidth}
    \vspace{0pt}

    \includegraphics[width=\linewidth]{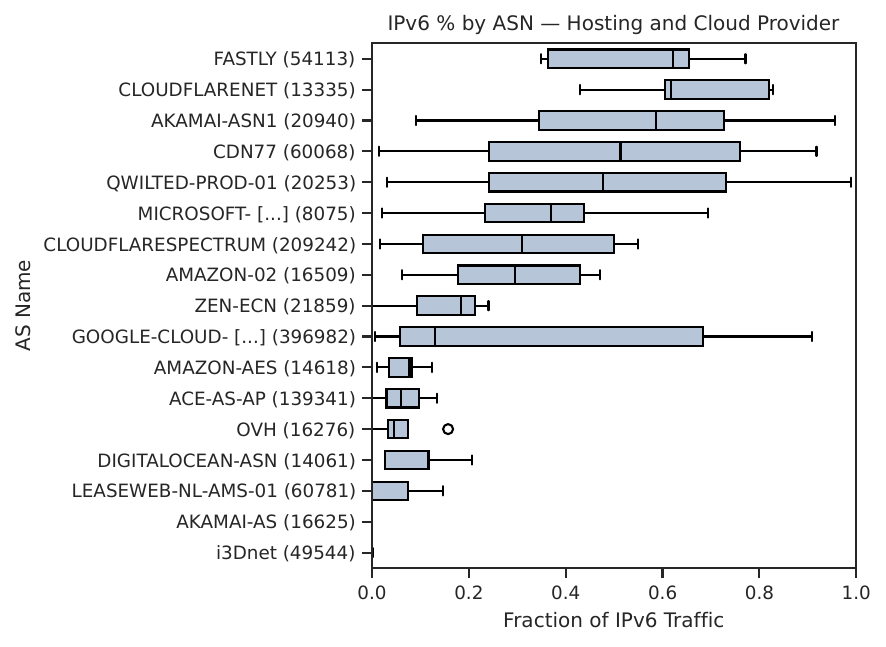}

    \includegraphics[width=\linewidth]{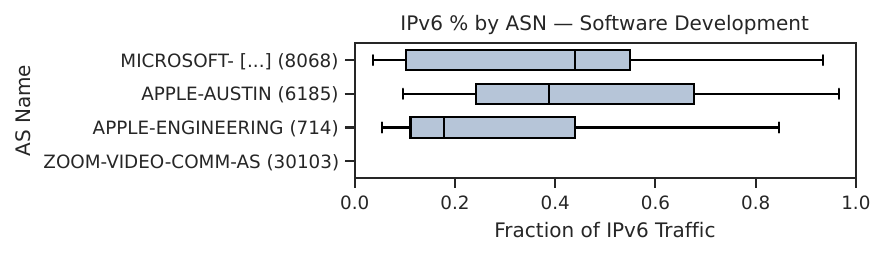}
  \end{minipage}
  \hfill
  \begin{minipage}[t]{0.48\linewidth}
    \vspace{0pt}

    \includegraphics[width=\linewidth]{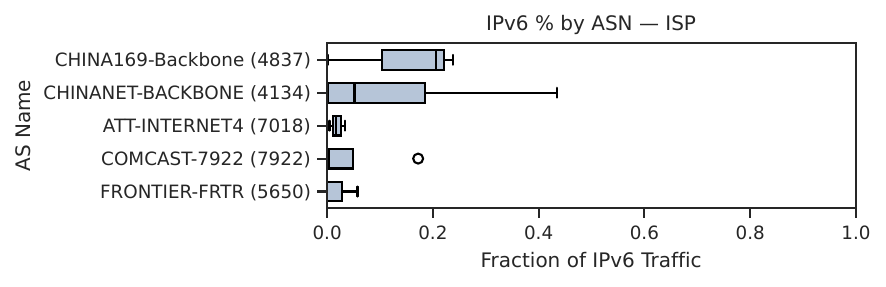}

    \vspace{8pt}

    \includegraphics[width=\linewidth]{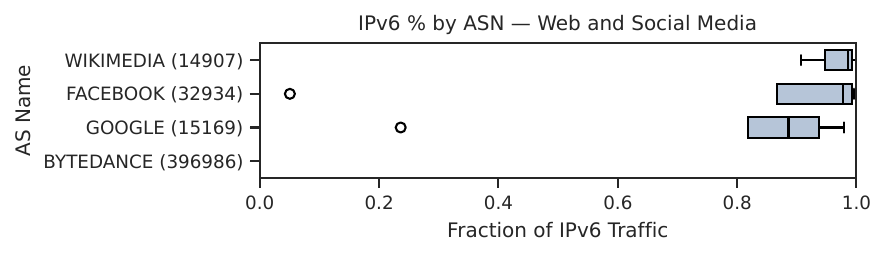}

    \vspace{8pt}

    \includegraphics[width=\linewidth]{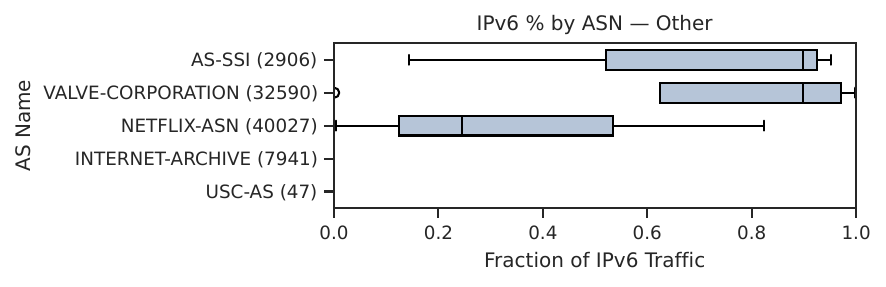}
  \end{minipage}

  \caption{
    Distribution of IPv6 byte fractions for 35 ASes observed at three or more locations, grouped by function.
    Within each category, ASes are sorted by median IPv6 byte fraction.
    Boxes show the interquartile range (25th-75th percentiles), and whiskers extend to 1.5$\times$IQR.
    Dots represent outliers.
  }
  \label{fig:as-ipv6-distributions}
\end{figure*}

In \autoref{sec:client/v6fraction}
  we saw that the fraction of IPv6 varies
  considerably by residence, and even by day within each residence.
We suggested that this variation is due to
  shifts in what specific services a user accesses each day.
To understand the role services play,
  we next look at which services lead and lag in IPv6 adoption?
Major services that lag in IPv6 support
  represent opportunities to improve deployment.
We focus on bytes rather than flows,
  as major services are more naturally identified by the volume of data they deliver
  than by the number of connections they initiate.

Here we identify services at the \emph{AS level}
  from their external IP address.
Our dataset includes the destination IP address of each flow, 
  but not website or service identities, 
  because of our privacy policy.
To bridge this gap, 
  we map the external IP address to its \emph{AS} from BGP routing tables,
  and to its \emph{domain} using reverse DNS.

\textbf{AS-Level Services:}
We identify service ASes from its external IP address.
To focus on the most significant contributors, 
  we consider only ASes whose traffic (in bytes)
  includes at least 0.01\% of aggregate volume at that residence.
We end up with 250 ASes or fewer per residence,
  and still over 97.4\% of all observed traffic.

We show the cumulative fraction of IPv6 bytes by AS in \autoref{fig:as-cdf}.
We observe that ASes with exclusively IPv4 traffic and exclusively IPv6 traffic both exist.
ASes with exclusively IPv4 traffic exist at all locations---at least one quarter of ASes at every location provide no IPv6 traffic.
Thus, some ASes prefer not to do IPv6, even when the client is dual-stack.
Conversely, not all locations have ASes with exclusively IPv6 traffic.
In particular, the highest IPv6 bytes fraction seen among ASes at \ResRouter{C} is 40\%.
We suspect this may be because some devices at \ResRouter{C} did not have IPv6 enabled, or had broken connectivity, and therefore even the ASes that prefer IPv6 could not use it.

To identify ASes with a consistent IPv6 (or IPv4) preference, we focus on 35 ASes observed at three or more residences.
We manually group them into five categories---Hosting, Software, ISP, Web and Social Media, and Other---based on primary function.
\autoref{fig:as-ipv6-distributions} shows the distribution of IPv6 byte fractions for these ASes, with ASes in each category sorted by median IPv6 use.
In the box plots, the boxes represent the interquartile range (25th to 75th percentiles), and whiskers extend to 1.5$\times$IQR.

We find that ISP ASes show consistently low IPv6 byte fractions, with medians at 20\% or lower, and no AS exceeding 50\%.
On the opposite end, Web/Social Media ASes (with the exception of ByteDance, the company behind Tiktok) show consistently high IPv6 byte fractions, with medians exceeding 90\%.

This AS-level analysis is helpful where it applies,
  but with many independent services hosted in shared clouds,
  AS-based analysis is limited.
We see a wide variation in IPv6 byte fractions for cloud services,
  even though clouds all offer IPv6 support.
Moreover, we know that some clouds
  lag in IPv6 adoption: OVH, DigitalOcean, and LeaseWeb show minimal IPv6 traffic, while i3D.net shows none.
But even for IPv6-ready clouds, the seeming contradiction between offering and use
  prompts us to study clouds separately in \autoref{sec:cloud}.

\textbf{Domain Level.}
We identify the domain associated with each flow via reverse DNS lookups on destination IP addresses.
We then aggregate traffic by domain,
  retaining only prominent domains:
  those that appear in three or more residences
  and generate at least 1\,GB of traffic.

We provide this long list of domains in \autoref{sec:appendix/client_service_details}.
We observe many domains with zero or near-zero IPv6 traffic across all locations---including services not captured at the AS level, such as GitHub, Twitch (\texttt{justin.tv}), and WordPress (\texttt{wp.com}).

For services that use clouds or CDNs, reverse DNS often maps to canonical names of the cloud rather than the actual service.
For example, a domain like \texttt{example.com} may resolve to \texttt{subdomain.cdn.net}, with reverse DNS revealing only the latter.
Consequently, domain-level aggregates for cloud and CDN platforms largely mirror AS-level patterns, with variations driven by the specific mix of tenant services accessed at each residence.
This limitation prompts our study of cloud-hosting in \autoref{sec:cloud}.

Finally, we note that Zoom, GitHub and \USC (a large U.S.~university) generate no IPv6 traffic.
This omission is significant: when key applications, workplaces, or campuses opt out of IPv6, it substantially limits users' IPv6 exposure.
We confirmed \USC's campus network lacks IPv6 support, meaning thousands of users are confined to IPv4.
This contributes to a negative feedback loop---services see little IPv6 demand and deprioritize support, reinforcing the university's decision.
Yet, this ``tragedy'' for IPv6 deployment also presents an opportunity: large campuses and enterprises can have a tangible impact on IPv6 adoption by enabling it, incentivizing services to adopt IPv6.

\subsection{Limitations in Analysis of Residences}
  \label{sec:client/limitations}

As with any real-world study, our study has limitations.

First, we study only five residences within a single metropolitan area, 
  with partial traffic visibility at two due to some individuals opting out.
With this small sample size, we make
  no claims that it represents the Internet or even users in our region.
However, 
  these five examples establish diversity of traffic mixes across residences,
  and show how IPv6 use is shaped by user choice of services
  and those services' support of IPv6.
We hope our findings motivate larger-scale study
  of how much IPv6 is in use by household.

Second, our dataset contains only flow-level information,
  and we do not have access to specific websites or traffic payloads.
This limitation was part of our IRB agreements
  to protect the privacy of our study participants.
We partially mitigate this limitation 
  by examining service ASes and reverse DNS names,
  but future studies may chose to evaluate services with 
  user consent and a more permissive IRB.

\section{Server-side Adoption of IPv6}
	\label{sec:serverside}

We now examine server-side IPv6 support.
While previous work examined main pages~\cite{nikkhah_assessing_2011,czyz_measuring_2014}
  and first-party resource~\cite{dhamdhere_measuring_2012} of
  sites drawn from top lists,
  our goal here is to deepen this analysis to consider
   \emph{third-party resources} and multiple pages at each website.
Our goal is to provide a more nuanced understanding of server-side IPv6 use,
  and to understand to what extent third-party content is a barrier to IPv6.

We consider third-party resources because
modern websites often depend on third-party components or CDNs
  to support social media (such as the Facebook ``like'' button or Disqus comment section),
  for user tracking and analytics, 
  for ad delivery, 
  and to support new fonts and JavaScript libraries.
These resources are essential to get a full picture of a website;
  recent studies of cookies found that including third-party resources
  discovered about $1.4\times$ more cookies than just examining the top page~\cite{urban_beyond_2020}.
While the website might be partially functional without
  some of these resources,
  we wish to evaluate the full website.
(Users may not miss third-party ads, but they are essential to the economics
  of many websites.)

\subsection{Measuring Servers}
	\label{sec:serverside/methodology}

We next describe our approach to evaluating server
  more completely than prior work.

First we require a list of potential websites.
We use the top 100k websites from the Tranco~\cite{LePochat2019} top 1M list dated 2024-10-16 (ID: \texttt{Z3Z3G}).
The Tranco list combines and averages website popularity rankings from multiple sources, including the Chrome User Experience Report (CrUX), Cloudflare Radar, Farsight, Majestic, and Cisco Umbrella, over a 30-day period;
  we discuss list trade-offs in \autoref{sec:serverside/limitations}.

We access websites using OpenWPM web privacy measurement framework~\cite{englehardt2016census}.
For each website, 
  OpenWPM spawns a Firefox browser
  to load and render the main page, including embedded resources.
For a more accurate measurement, as suggested in prior work~\cite{urban_beyond_2020}, it then follows five randomly chosen links (an arbitrary number; fewer if the page contains too few links),
  selected to be in the same domain.
Specifically, we ensure that the links do not lead us outside the \emph{eTLD+1} of the crawled website,
  that is, a domain name consisting of one label and a public suffix
  as defined by the Public Suffix List~\cite{noauthor_privacy_2021,Mozilla07a}.
Some websites attempt to detect bots,
  so OpenWPM uses standard cloaking techniques such as simulating mouse movements, scrolling through the page, and pausing for a few seconds before
  simulating a click to navigate to the next page.
For each top site, OpenWPM records the HTTP requests made by the browser, the DNS query results for each request, the IP address used to establish the connection, as well as the HTTP redirects and responses from the server(s).
Measurements were conducted on machines running OpenSUSE Tumbleweed and Fedora, connected to residential and academic networks respectively, both with full IPv4 and IPv6 support.

We conduct measurements in October 2024, 
  then repeat them in April and July 2025
  to evaluate how the results evolve over the nine-month interval.

\subsection{To what extent do websites support IPv6?}
	\label{sec:serverside/topsites}

\begin{figure*}
  \centering
  \begin{subfigure}[b]{\textwidth}
      \centering
      \resizebox{0.85\textwidth}{!}{%
        \begin{tabular}{l r r r r r r r}
          \hline
          \textbf{Category} 
          & \multicolumn{2}{c}{\textbf{Oct 2024}} 
          & \multicolumn{2}{c}{\textbf{Apr 2025}} 
          & \multicolumn{2}{c}{\textbf{Jul 2025}} 
          & \textbf{Change} \\
          & \textbf{Count} & \textbf{(\%)} 
          & \textbf{Count} & \textbf{(\%)} 
          & \textbf{Count} & \textbf{(\%)} 
          & \textbf{(\%)} \\
          \hline
          Total & 100\,000 & -- & 100\,000 & -- & 100\,000 & -- & -- \\
          \hspace{0.3cm} Loading-Failure (\texttt{NXDOMAIN}) & 12\,355 & -- & 13\,107 & -- & 13\,376 & -- & -- \\
          \hspace{0.3cm} Loading-Failure (Others) & 4\,457 & -- & 4\,567 & -- & 4\,802 & -- & -- \\
          \hline
          \hspace{0.3cm} \textbf{Connection Success} 
            & \textbf{83\,188} & \textbf{(100\%)} 
            & \textbf{82\,326} & \textbf{(100\%)} 
            & \textbf{81\,822} & \textbf{(100\%)} 
            & \textbf{(0.0)} \\
          \hspace{0.6cm} Unknown Primary Domain 
            & 8 & (0.0\%) & 6 & (0.0\%) & 3 & (0.0\%) & (0.0) \\
          \hspace{0.6cm} IPv4-only (A-only domain) 
            & 48\,434 & (58.2\%) & 47\,388 & (57.6\%) & 47\,158 & (57.6\%) & (-0.6) \\
          \hspace{0.6cm} \textbf{AAAA-enabled Domain} 
            & \textbf{34\,746} & \textbf{(41.8\%)} & \textbf{34\,932} & \textbf{(42.4\%)} & \textbf{34\,661} & \textbf{(42.4\%)} & \textbf{(0.6)} \\
          \hspace{0.9cm} IPv6-partial (some A-only resources) 
            & 24\,779 & (29.8\%) & 24\,624 & (29.9\%) & 24\,384 & (29.8\%) & (0.0) \\
          \hspace{0.9cm} \textbf{IPv6-full (AAAA for all resources)} 
            & \textbf{9\,967} & \textbf{(12.0\%)} & \textbf{10\,308} & \textbf{(12.5\%)} & \textbf{10\,277} & \textbf{(12.6\%)} & \textbf{(0.6)} \\
          \hspace{1.2cm} Browser Used IPv4 
            & 1\,122 & (1.4\%) & 870 & (1.0\%) & 1\,189 & (1.5\%) & (0.1) \\
          \hspace{1.2cm} \textbf{Browser Used IPv6 Only} 
            & \textbf{8\,845} & \textbf{(10.6\%)} & \textbf{9\,438} & \textbf{(11.5\%)} & \textbf{9\,088} & \textbf{(11.1\%)} & \textbf{(0.5)} \\
          \hline
        \end{tabular}
      }
      \caption{Tabular Representation comparing Oct 2024, Apr 2025 and Jul 2025 measurements}
  \end{subfigure}
  \begin{subfigure}{\textwidth}
      \centering
      \includegraphics[width=0.65\textwidth,trim=0.5cm 3cm 11cm 0cm, clip]{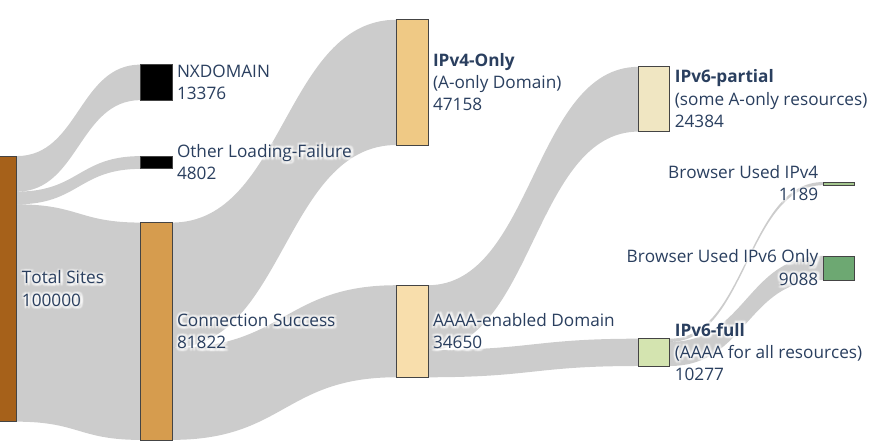}
      \caption{Sankey Diagram of IPv6-readiness based on Jul 2025 measurements}
  \end{subfigure}
  \caption{
    Sankey diagram and tabular breakdown of the Tranco top 100,000 websites, classified into IPv4-only, IPv6-partial, or IPv6-full.
  }
  \label{fig:serverside-results}
\end{figure*}

\begin{figure}
  \centering
  \includegraphics[width=0.9\linewidth]{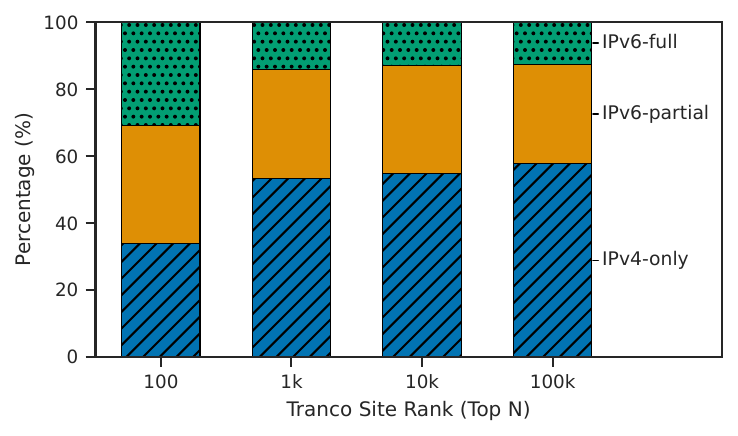}
  \caption{
    Stacked bar chart comparing IPv6 readiness of top N websites (N=100, 1k, 10k, 100k) from Tranco top 1M list.
  }
  \label{fig:serverside-top-n-comparison}
\end{figure}

Using the dataset we collected, we begin to address the question \emph{To what extent does each website support IPv6?}
Rather than categorizing top sites simply into a binary of \emph{fully loadable} or \emph{not fully loadable} over IPv6, we examine the individual steps involved in accessing a website on a machine with dual-stack connectivity and identify specific factors that may prevent a site from being fully accessible over IPv6.
We then provide a detailed classification of top sites based on these factors.

\textbf{Classifying Degrees of IPv6 Support.}
We convey the range of levels of IPv6 support as IPv4-only, IPv6-partial, IPv6-full, or loading-failure.

We classify websites lacking a AAAA DNS record as \emph{IPv4-only}.
We recognize that some resources may be accessible via IPv6,
  but without the main page, that matters little.

A website is \emph{IPv6-partial}
  when the main page is available via IPv6,
  but some resources are IPv4 only.
An IPv6-only user will likely see a degraded website for such pages.
For these websites, we report the fraction of resources that are IPv6-capable.

A website is considered \emph{IPv6-full} if both its main page and all embedded resources are accessible via IPv6.

Finally, we label websites \emph{loading-failure}
  for pages with DNS or HTTP errors and
  TLS failures.
Resources that face such failure are excluded from our analysis, as these failures are independent of IP version and thus are orthogonal to our discussion.

\textbf{Evaluation Methodology.}
Two factors influence how much of a page we see as IPv6-ready.

First, the Happy Eyeballs algorithm requires web browsers
  to query IPv6 and IPv4 in parallel, 
  with slight delays introduced to prioritize IPv6~\cite{schinazi_happy_2017}.
Because both protocols are queried simultaneously, 
  it may change which protocol actually retrieves the content.
However, we check IPv6 availability, rather than whether IPv6 is ultimately used,
  so our results are not affected by which protocol wins the happy eyeballs ``race''.

Second, retrieving the main page and each resource involves multiple steps, 
  any of which can fail.
If the main page fails, the entire website is classified as a loading-failure.
A full retrieval of each resource (the main page or an embedded resource) requires:
  a successful DNS query,
  TCP connection establishment,
  successful TLS negotiation (for HTTPS),
  an HTTP request,
  and finally the retrieval of the complete resource.
Additionally, any resource can trigger one or more HTTP redirects.
We follow all redirects whenever possible, 
  and report the outcome for the final page in the redirect chain.

\textbf{Results.}
We present the results of our analysis of Tranco top 100,000 websites, measured between October 2024 and July 2025, in \autoref{fig:serverside-results}.
Our discussion focuses on the July 2025 measurements, with comparisons to previous measurements to highlight changes over time.
We ignore the 18,178 websites that fail to connect
  and examine the remaining 81,822.

First, we see that the majority of websites (47,158 or 57.6\% of reachable sites)
  are IPv4-only.
While this fraction is significant progress
  from the 96\% of top-10k websites that were IPv4-only
  in 2014~\cite{czyz_measuring_2014},
  it is disappointing that, ten years later, 
  more than half of the top-100k websites \emph{still} lack IPv6 access.

Second,
  29.8\% of the reachable websites (24,384)
  are IPv6-partial.
These web pages access some resources that are IPv4-only.
Although the webpage author started to support IPv6 with the main page,
  as they drew in additional content, 
  they were not careful to make sure their providers were as diligent.

Finally, 12.6\% of sites (10,277) are IPv6-full.
Although here, we see that only 11.1\% (9,088)
  \emph{actually} used IPv6 for all parts.
We believe this final drop-off,
  about 1 in 10 times of pages that could be IPv6,
  occurs because 
  IPv4 sometimes wins the Happy Eyeballs race.

Comparing these results with October 2024 reveals a slight but consistent increase in IPv6 adoption: the share of IPv4-only websites decreased by 0.6\%, while the proportion of IPv6-full websites increased by the same amount.
Although modest, these changes reflect ongoing progress toward broader IPv6 deployment.

To examine how website popularity affects IPv6 readiness, \autoref{fig:serverside-top-n-comparison} compares IPv6 readiness for the top 100, 1k, 10k, and 100k sites.
The top 100 sites stand out, with 30.1\% being IPv6-full---more than double the rate observed across the top 100k (12.6\%).
This analysis suggests the \emph{most} popular websites have invested in IPv6,
  but the ``long tail'' of moderately popular websites lags in IPv6 adoption.

We also find that \emph{not} clicking the five internal links (considering only the main page and its embedded resources) raises the fraction of IPv6-full websites from 12.5\% to 14.1\%.
This 1.6\% jump is over twice the \emph{actual} 0.6\% growth seen in nine months, demonstrating the need to look beyond the main page.

\subsection{Which Resources Make Websites IPv6-partial?}
  \label{sec:serverside/resource-domains}

\begin{figure}
  \centering
  \includegraphics[width=0.9\linewidth]{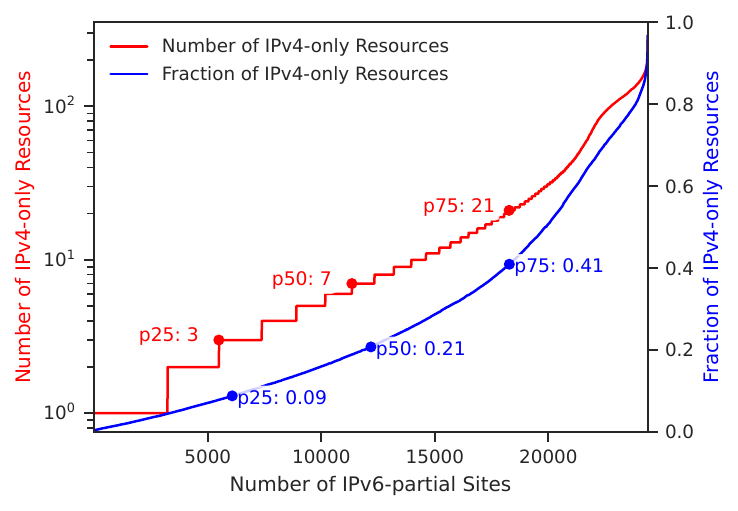}
  \caption{
    CDF illustrating the total count (red, left axis) and fraction (blue, right axis) of IPv4-only resources used by IPv6-partial websites.
  }
  \label{fig:serverside-ipv6-partial-cdf}
\end{figure}

\begin{figure}
  \centering
  \includegraphics[width=0.9\linewidth]{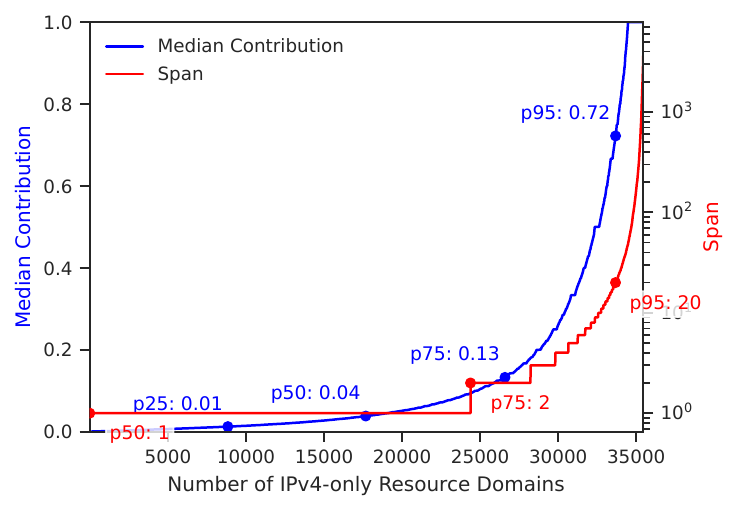}
  \caption{
    CDFs of median contribution (blue, left axis) and span (red, right axis) for IPv4-only domains used by IPv6-partial websites.
  }
  \label{fig:serverside-median-contribution-span-cdf}
\end{figure}

\begin{figure}
  \centering
  \includegraphics[width=0.9\linewidth]{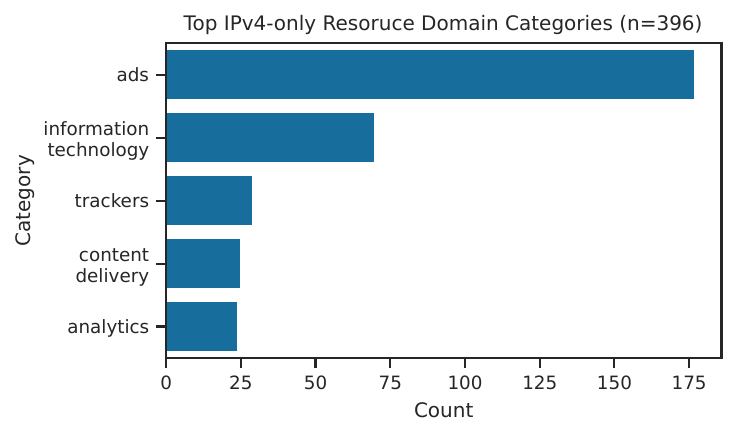}
  \caption{
    Most common categories among the 396 IPv4-only resources with high span ($\geq$ 100 websites), based on VirusTotal~\cite{noauthor_virustotal_nodate} domain categorization.
  }
  \label{fig:serverside-v4only-resource-domain-categories}
\end{figure}

\begin{figure}
  \centering
  \includegraphics[width=0.85\linewidth]{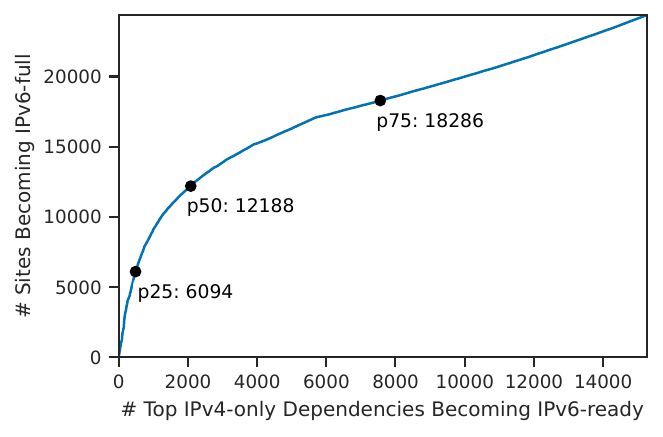}
  \caption{
    CDF showing how many IPv6-partial top websites become IPv6-full as IPv4-only resources adopt IPv6 in descending order of their span.
  }
  \label{fig:serverside-third-party-cdf}
\end{figure}

Given that two-thirds of IPv6-enabled websites are
  only IPv6-partial raises the question: 
  \emph{Which resources are preventing IPv6-full?}
Resources lacking IPv6 support
  are ``low-hanging fruit'' to 
  transition more websites to IPv6-full.

To understand how much change is needed to improve IPv6-partial websites,
  \autoref{fig:serverside-ipv6-partial-cdf} 
  shows the number (red line and axis)
  and fraction (blue line and axis)
  of IPv4-only resources seen in IPv6-partial sites.
We see that almost all IPv6-partial sites
  depend on multiple IPv4-only resources (75\% of sites need three or more),
  even though the majority of resources are usually available on 
  IPv6 (from the blue curve, 75\% of IPv6-partial sites have more 
  than half the resources available via IPv6).
Although most resources are IPv6-ready, the fact that most IPv6-partial sites depend on multiple IPv4-only domains means that widespread improvements are needed to achieve full adoption.

To distinguish same-origin from cross-origin dependencies,
  we classify IPv4-only resources as either \emph{first-party} or \emph{third-party} by comparing their eTLD+1 domains
  to that of the crawled site; a mismatch indicates a third-party domain.
Surprisingly, 565 out of 24,384 IPv6-partial websites (2.3\%)
  are IPv6-partial because of IPv4-only first-party domains---presumably
  something the website can correct if they choose to do so,
  since they already operate IPv6-enabled servers.
For example, IPv6-full \url{www.national-geographic.org} uses
  images from IPv4-only \url{assets.national-geographic.org}.
While rare, these cases should be easy to correct.

We assess the impact of IPv4-only domains using
  \emph{span} and \emph{median contribution} from prior work~\cite{bajpai_longitudinal_2019}.
Span of an IPv4-only domain
  counts how many IPv6-partial websites depend on it,
  and median contribution is the median of the fraction of IPv4-only resources
  it supplies to all websites that depend on it.
Larger values for either shows a more influential resource
  holding back more IPv6-partial websites than other resources.

\autoref{fig:serverside-median-contribution-span-cdf} gives CDFs
  of span and median contribution for all IPv4-only eTLD+1 domains used by IPv6-partial websites.
The span distribution is highly skewed, with a long tail: 
  while most IPv4-only resources are used by only one or two other sites,
  a few are very widely used.
Specifically, 75\% of these domains are used by at most two websites, 
  and for 75\% of them, the median 
  fraction of IPv4-only resources they provide to each dependent website is no more than 13\%.
In contrast, a small number of domains exert a very high influence.
At the 95th percentile, an IPv4-only domain appears on 20 websites and contributes, on median, 72\% of the IPv4-only resources for each of those websites.
A handful of such domains are used by over 1,000 websites, and some are responsible for all IPv4-only resources on the websites that depend on them.

\textbf{Heavy-hitter IPv4-only resources.}
We now focus on the ``heavy-hitters''---396 IPv4-only resources with a span of at least 100 websites.
What functionality do these domains provide?
We answer this question two different ways.

First, we characterize these IPv4-only resources with VirusTotal's domain categorization~\cite{noauthor_virustotal_nodate}.
\autoref{fig:serverside-v4only-resource-domain-categories} shows the most common categories among these high-impact domains.
Advertising is most frequent, accounting for nearly half of the domains.
While we expect many third-party 
  advertising services, 
  we are surprised so many remain IPv4-only.
Advertising seeks to maximize reach and 
  IPv6 may enhance their ability to track users
  with per-user IPv6 addresses instead of shared, NAT'ed IPv4.
Other frequently observed categories include tracking, CDN and analytics.

Next, we examine the resource types served by prominent IPv4-only resources through the heatmap shown in \autoref{sec:appendix/serverside-third-party-laggards} (in the appendix for space).
Images are the most frequently served resource type, followed by \texttt{sub\_frame}, \texttt{xmlhttprequest}, and JavaScript.
Thus, IPv6-only users may encounter broken images or impaired functionality on websites that depend on these domains.

To identify specific third-party domains that would have large impact if they adopted IPv6, we simulate a scenario where IPv4-only third-party domains enable IPv6 one at a time, in descending order of their span.
At each step, we compute the fraction of currently IPv6-partial websites that would transition to IPv6-full.
The resulting CDF is shown in \autoref{fig:serverside-third-party-cdf}.
The distribution exhibits a long tail: enabling IPv6 on just the top 500 (3.3\%) IPv4-only third-party domains would allow over 25\% of IPv6-partial websites to become IPv6-full.
However, for all 24,384 partially IPv6-partial websites to become IPv6-full, over 15,000 third-party domains would need to enable IPv6.
Thus, prioritizing IPv6 adoption among a relatively small set of high-impact third-party domains can substantially increase the number of IPv6-full websites, but achieving universal IPv6 readiness will require a much broader effort.

\subsection{Limitations in Service Analysis}
  \label{sec:serverside/limitations}

Our analysis inherits the known limitations of top-list based studies---results may vary by choice of list, its age, and its size~\cite{ruth_toppling_2022,Scheitle18a}.
We choose the Tranco list for its use of multiple data sources and its wide adoption in both academic and non-academic research.
While no list perfectly reflects popularity, we believe Tranco provides a plausibly representative sample for our study.

While we observed no significant differences between our two vantage points in the Western U.S.~(one residential, one academic), 
  results may vary by location
  due to variations in CDN and advertising selection.
Nonetheless, we believe our findings broadly reflect the current state of IPv6 adoption among popular websites.

A further limitation arises from the possibility that dual-stacked sites 
  load different embedded resources depending on which protocol 
  wins Happy Eyeballs.
This behavior can lead to misclassification, 
  since resources selected in a page can vary by the protocol used to access it.
An otherwise IPv6-full site may appear to be IPv6-partial 
  if it is loaded by IPv4 and dynamically selects IPv4-only resources.
To estimate how often such misclassifications happen, we examined IPv6-partial sites where all IPv4-only resource FQDNs contained substrings like \texttt{v4}, \texttt{ipv4}, or \texttt{px4}, suggesting intentional version-specific subdomain use.
We found 106 potentially misclassified sites (0.4\% of all IPv6-partial sites), indicating this rare edge case is unlikely to impact our overall conclusions.

Lastly, our treatment of all third-party resources as a single category may overlook differences in functional importance.
In practice, missing resources vary in impact---a broken ad is less disruptive than a missing JavaScript library.
Nonetheless, since ads and trackers often play a central role in monetizing websites, their IPv6 availability remains important to examine.

\section{Cloud Adoption of IPv6}
  \label{sec:cloud}

With over half of all web traffic served via CDNs~\cite{sumits_internet_2017} 
  and 90\% of companies using cloud computing~\cite{krivec_cloud_2024},
  cloud and CDN support for IPv6 is essential.
Today, all cloud and CDN platforms support IPv6,
  so one would think that customers can just ``push a button'' to enable IPv6 
  in their cloud servers.
However, as shown in \autoref{fig:as-ipv6-distributions}, cloud and CDN traffic is a mix of IPv4 and IPv6, with the proportions varying widely across locations for many providers.

This section explores
  cloud and CDN (hereafter, just ``cloud'') support for IPv6 and why
  it is more difficult than one might expect for a customer to turn it on.
We use the term ``cloud'' for both clouds and CDNs for simplicity,
  and because modern clouds and CDNs increasingly overlap,
  with traditional clouds providing CDNs (such as Amazon CloudFront)
  and vice versa (Akamai Compute).
    
\subsection{Are Some Clouds IPv6-Heavy?}
  \label{sec:cloud/cloud-share}

\begin{figure*}
  \centering
  \includegraphics[width=0.8\linewidth]{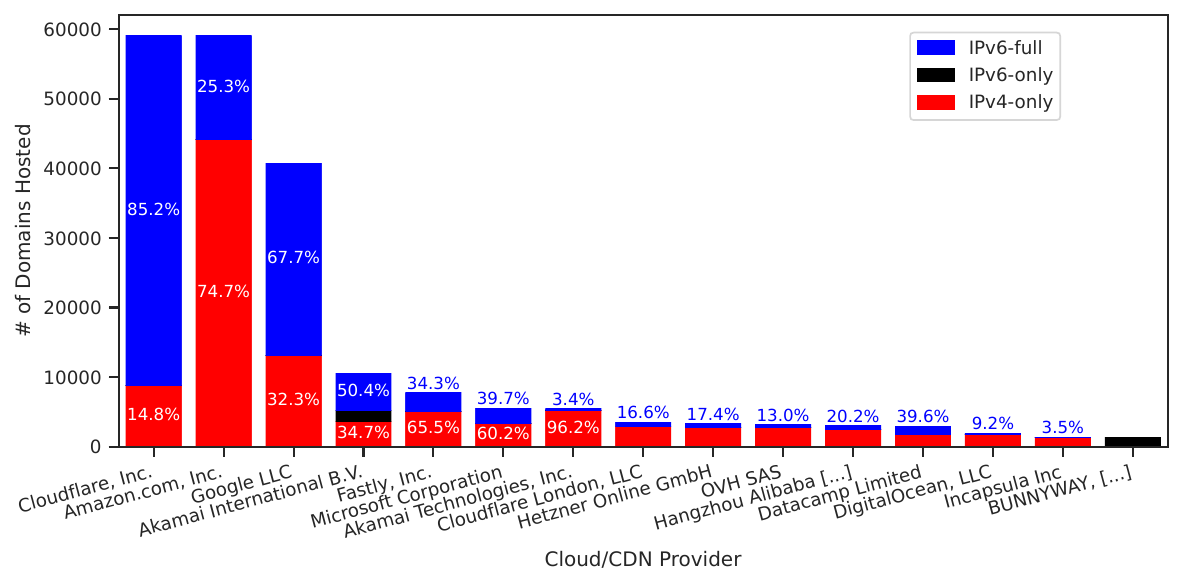}
  \caption{
    IPv6-readiness breakdown of the top 15 cloud providers by number of domains.
  }
	\label{fig:cloud-domain-breakdown}
\end{figure*}

We first ask: are some clouds IPv6-heavy?
That is, do some clouds or CDNs show more (or less) IPv6 traffic than typical?
We assume IPv6 support is similar across all clouds,
  so differences in IPv6 use suggest something different about the clouds,
  and a possibility to improve on or emulate what they are doing.

We examine IPv6-use by cloud
  by comparing the proportion of IPv6-full domains
  hosted by each provider against those of others.
We reuse the server-side data from \autoref{sec:serverside/methodology},
  beginning with the Tranco top 100k list, crawling each site, 
  and simulating up to five random link clicks.
For each page load, we record the fully qualified domain names (FQDNs) of all loaded resources, resulting in a dataset of 265,248 FQDNs.
With a large list of popular websites, we expect it to include most major clouds as well as their customers.

As before, we classify domains as IPv4-only (A record only) or IPv6-full (both A and AAAA).
Unlike top sites, we observe some domains that only have AAAA records;
  we classify AAAA-only domains as \emph{IPv6-only}.

We identify a domain's cloud provider by the AS that originates the BGP prefix containing the domain's IP address.
Because a single cloud may operate multiple ASes, we map each AS number to an organization name using CAIDA's AS-to-Organization dataset~\cite{caida-as2org}.

\autoref{fig:cloud-domain-breakdown} compares the top 15 clouds
   ranked by the number of domains they host, 
  and their degree of IPv6 readiness.
We provide the full breakdown in \autoref{sec:appendix/cloud-domain-breakdown} in the appendix.
Collectively, these clouds account for 76\% of all domains in our dataset.

While all clouds support IPv6 and so all cloud customers \emph{could} use it,
  we see that \emph{only 3\% to 85\% of cloud customers actually chose to support IPv6}.
For some reason there is a huge variation in practice, which we explore further in \autoref{sec:cloud/cloud-obstacles}.
Cloudflare, Amazon, and Google collectively host nearly 60\% of the domains we analyze.
Among them, Cloudflare (85.2\%) and Google (67.7\%) demonstrate strong IPv6 support, whereas Amazon trails considerably at 24.6\%.

Smaller clouds tend to have lower IPv6 adoption rates, 
  except for Bunnyway.
The Bunnyway CDN \url{bunny.net} is almost entirely IPv6-only (99.5\%),
  with 1,203 domains with IPv6 addresses (AAAA records) at
  bunny.net IP addresses,
  but IPv4 (A records) hosted by Datacamp Limited, operator of CDN77.
This relationship is a partnership between the two companies,
  with Bunnyway using Datacamp servers~\cite{Datapacket25a},
  an organizational relationship not captured by the AS-to-Org dataset we used.
A similar attribution issue arises with Akamai, where 1,570 IPv6-only domains appear to be hosted by Akamai International B.V., while their IPv4 addresses are associated with Akamai Technologies, Inc.
In this case, the distinction is between two entities within the same company, but again, the AS-to-Org dataset treats them as separate organizations.

\subsection{Do Some Clouds Make IPv6 Harder?}
  \label{sec:cloud/cloud-obstacles}

\begin{figure*}
  \centering
  \includegraphics[width=\linewidth]{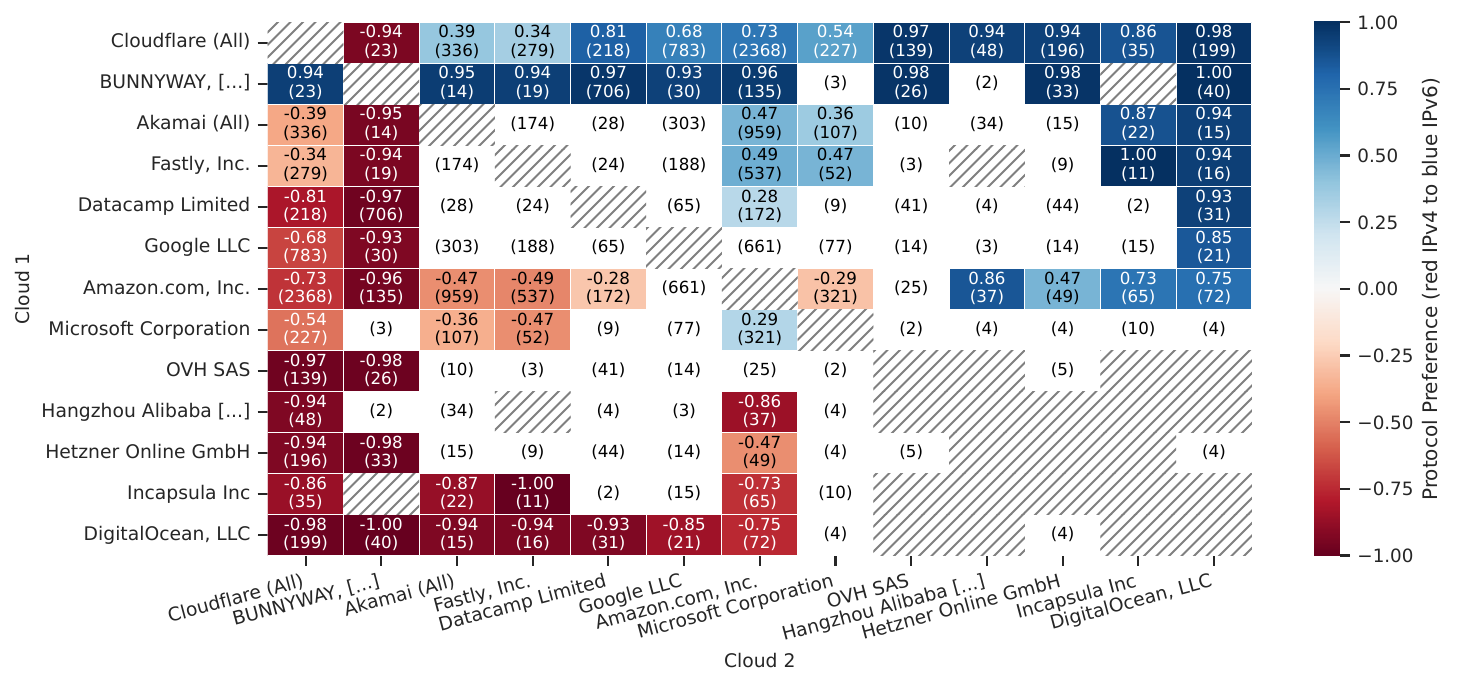}
  \caption{
    Heatmap showing the results of comparing IPv6 support of pairs of cloud providers using two-sided Wilcoxon signed-rank tests.
  }
  \label{fig:cloud-multi-provider-wilcoxon}
\end{figure*}

What accounts for the significant variation in IPv6 use we see in \autoref{sec:cloud/cloud-share}?
While all clouds support IPv6,
  we hypothesize that some make its use harder or easier.
For example, some enable it by default,
  while others require one explicitly enable it on a control setting.

However, comparing IPv6 usage by tenant across clouds
  is complicated because our data combines two factors:
  the cloud's support for IPv6,
  and the tenant's interest in IPv6.
To disambiguate these two factors,
  we consider tenants that use multiple clouds,
  since that should hold the interest of the tenant constant
  and emphasize any differences in cloud mechanisms.

To evaluate tenant behavior,
  we identify \emph{multi-cloud tenants}
  as those tenants that use at least two different clouds.
Concretely, we define a multi-cloud tenant as any eTLD+1 domain (see \autoref{sec:serverside/methodology} for definition)
  whose subdomains are hosted by two or more different organizations.
To illustrate, \texttt{apnic.net} is a multi-cloud tenant
  with a number of subdomains, including 
  \texttt{www}, \texttt{blog}, \texttt{cgi1}, \texttt{webdist.data},
  \texttt{info}, and \texttt{login}.
Of these, most are hosted by Cloudflare and are IPv6-full,
  but the last two are hosted by Amazon and are IPv4-only.
Our expanded Tranco dataset identifies 21,314 multi-cloud tenants.

We then compare a tenant's IPv6-readiness across the multiple clouds that serve it.
For each tenant we first compute the fraction of subdomains that are IPv6-full under each cloud.
(For example, in a comparison of Cloudflare and Amazon, \texttt{apnic.net} would count as 1.0 for Cloudflare and 0.0 for Amazon.)
Then we assess differences between pairs of clouds
  by applying a two-sided Wilcoxon signed-rank test.
This test requires at least two shared tenants where the two clouds differ in IPv6 support.
In our dataset, 67 of the 78 possible cloud pairs meet this criterion.

To quantify the magnitude of these differences, we calculate the effect size $r \in [-1,1]$ for each comparison
  where $r$ uses the Wilcoxon signed-rank test.
A positive effect size ($r>0$) indicates that, for shared tenants, subdomains hosted by the first cloud are more often IPv6-full than those hosted by the second, with $r=1$ indicating that the first cloud \emph{always} wins.
Negative effect sizes indicate the opposite.
Finally, we correct for multiple comparisons using the Holm-Bonferroni method (with significance level $\alpha = 0.05$) to control the family-wise error rate~\cite{Holm-Bonferroni}.

\autoref{fig:cloud-multi-provider-wilcoxon} shows a heatmap
  comparing clouds by effect size $r$.
Each row and column is one of the 13 clouds,
  ordered here by how often the cloud has greater IPv6 support than the others by $r$ value.
Each cell is colored by size of the difference ($r$),
  with blue showing better IPv6 coverage and red, lower.
White cells have statistically indistinguishable IPv6 coverage,
  and hatched regions are not comparable because of insufficient examples.
Parentheses show the number of tenants where the two clouds differ in IPv6 support.
(The table is symmetric.)

We see large differences between several providers.
Cloudflare and Akamai, two large providers, both have better than typical
  IPv6 support.
Bunnyway also stands out, but this difference is largely because
  of their shared hosting with Datacamp as described in~\autoref{sec:cloud/cloud-share}.
Several other large providers (Google, Amazon, and Microsoft)
  show typical performance,
  while smaller providers seem to show poorer-than-average IPv6 support (consistent with our findings in~\autoref{sec:client/leaders-laggards}).
CDN-first providers generally show stronger IPv6 adoption, while traditional cloud providers tend to rank lower.

We attribute these differences to cloud deployment policies.
Cloudflare and Akamai, the providers with greatest IPv6 support, 
  employ a \emph{default IPv6-on} policy since 
  2014~\cite{levy_cloudflare_ipv6_default_2014}
  and 2016~\cite{nygren_akamai_2018}, respectively.
By contrast, the large cloud providers---Amazon, Google, and Microsoft---offer inconsistent IPv6 support across their products: while Amazon CloudFront and Azure CDN are IPv6-on-by-default~\cite{cloudfront_announcing_2016,azurecdn_enable_2021}, other services remain IPv6-opt-in, or lack support entirely~\cite{aws_ipv6_2025,azure_overview_2024,gcp_ipv6_2025}.

\subsection{Which Clouds Lead and Lag in IPv6?}
  \label{sec:cloud/cloud-service-leaders-laggards}

\begin{table*}
  \centering
  \resizebox{0.85\textwidth}{!}{
    \begin{tabular}{@{}l >{\raggedright\arraybackslash}p{0.25\linewidth} l r r r@{}}
      \toprule
      \textbf{Cloud Provider} & \textbf{Service} & \textbf{IPv6 Support} & \textbf{\# IPv6-ready} & \textbf{\# Total} & \textbf{\% IPv6-ready} \\
      \midrule

      \multirow{1}{*}{Cloudflare}
        & Cloudflare CDN                   & Default-On, Opt-out~\cite{cloudflare_ipv6_2025}  & 3\,086 & 4\,402 & 70.1\% \\
      \midrule

      \multirow{1}{*}{Bunny.net}
        & bunny.net CDN                    & Default-On~\cite{dejan_grofelnik_pelzel_return_2019}  & 1\,003 & 1\,004 & 99.9\% \\
      \midrule

      \multirow{2}{*}{Akamai}
        & Akamai CDN                       & Default-On, Opt-out~\cite{yeung_akamai_2023}  & 3\,620 & 7\,419 & 48.8\% \\
        & Akamai NetStorage                & Default-On, Opt-out~\cite{yeung_akamai_2023}  &   791  & 1\,633 & 48.4\% \\
      \midrule

      \multirow{2}{*}{DataCamp}
        & CDN77                            & Yes~\cite{cdn77enablesipv6_616404_2017}  &   673  &   759  & 88.7\% \\
        & bunny.net CDN                    & Default-On~\cite{dejan_grofelnik_pelzel_return_2019}  &   217  & 1\,300 & 16.7\% \\
      \midrule

      \multirow{3}{*}{Google}
        & Google Cloud Run                 & Yes~\cite{gcp_ipv6_2025}  &   334  &   334  & 100.0\% \\
        & Google App Engine                & Default-On~\cite{googleappengineipv6}  &   150  &   150  & 100.0\% \\
      \midrule

      \multirow{7}{*}{Amazon}
        & Amazon CloudFront CDN            & Default-On, Opt-out~\cite{cloudfront_announcing_2016,cloudfrontipv6toggle}  &  9\,142& 12\,851& 71.1\% \\
        & Amazon Elastic Load Balancer     & Partial~\cite{aws_ipv6_2025}  &    201 &  2\,731&  7.4\% \\
        & Amazon Global Accelerator        & Yes~\cite{aws_ipv6_2025}  &      4 &    150 &  2.7\% \\
        & Amazon S3                        & Yes~\cite{aws_ipv6_2025}  &      7 &  1\,862&  0.4\% \\
        & Amazon API Gateway               & Yes~\cite{aws_ipv6_2025}  &      0 &    419 &  0.0\% \\
        & Amazon Web App. Firewall  & Yes~\cite{aws_ipv6_2025}  &      0 &    134 &  0.0\% \\
      \midrule

      \multirow{6}{*}{Microsoft}
        & Azure Stack/IoT Edge             & Yes  &  1\,134&  1\,134& 100.0\% \\
        & Azure Front Door CDN             & Always On~\cite{microsoft_forwarding_2024}  &    913 &    913 & 100.0\% \\
        & Azure Cloud Services / VMs       & Yes~\cite{azure_overview_2024}  &      2 &    607 &  0.3\% \\
        & Azure Websites                   & Unknown  &      0 &    544 &  0.0\% \\
        & Azure Blob Storage               & Unknown  &      0 &    354 &  0.0\% \\
      \bottomrule
    \end{tabular}
  }
  \caption{
    IPv6 adoption across services for selected cloud providers.
    Within each provider, services are sorted by the percentage of IPv6-ready domains.
  }
  \label{tab:cloud-services}
\end{table*}

Which cloud services lead or lag in IPv6 adoption, and why?
Clouds offer a range of services, and even within a single provider, IPv6 adoption can vary by service.
This variance may depend on factors such as when IPv6 support was introduced by the provider, and how easily it can be enabled.
Understanding these patterns can help explain what drives or impedes adoption.

We identify cloud services by domain names.
Following work by He et al.~\cite{he_next_2013}, 
  we use CNAME records in cloud deployments to infer the identities of services.
We start with our earlier list of 265k FQDNs and resolve the DNS records for each through CNAMEs or chains of CNAMEs, resulting in 430,924 unique domains.
We then extract domain suffixes covering 100 or more FQDNs (such as \texttt{*.s3.amazonaws.com}) and manually map them to cloud services using official documentation.
\autoref{tab:cloud-services} lists the 20 identified services, 
  their 7 providers, and corresponding IPv6 adoption rates.

We first ask, how are the services with default IPv6-on policies performing?
Cloudflare, Akamai and Amazon CloudFront CDNs have had default IPv6-on policies for nearly a decade~\cite{nygren_akamai_2018,levy_cloudflare_ipv6_default_2014,cloudfront_announcing_2016}, yet adoption remains incomplete, with 30-50\% of domains still IPv4-only.
These platforms allow tenants to opt out of IPv6, and many have done so.
In contrast, Azure Front Door does not permit IPv6 to be disabled~\cite{microsoft_forwarding_2024}, and all its domains are IPv6-ready.
This suggests that default-on policies alone are insufficient---tenants often disable IPv6 to avoid perceived complexity.

Although Amazon announced IPv6 support for S3 and CloudFront in 2016~\cite{cloudfront_announcing_2016}, S3's actual adoption is near zero (0.4\%)!
While CloudFront enables IPv6 transparently,
  S3 requires tenants to enable IPv6 by changing to
  a different S3 URL.  %
Since many tenant pages embed S3 URLs, this requirement forces tenants
  to modify each page.
With near-zero adoption after nine years, this form of
  ``opt-in by code change'' slows deployment.
Elastic Load Balancers also require a CNAME update to enable IPv6~\cite{elb_configure_ipv6}, but since ELB domains are rarely hardcoded, adoption is slightly higher at 7.4\%.

These observations prompt us to recommend that all services should transparently add IPv6, adding AAAA records rather than requiring domain changes.
Disabling IPv6 should be permitted only in cases of service-breaking issues.
This default-on, no-disable approach achieves usage rates of 100\%, instead of less than 10\% or 1\% when manual opt-in and code changes are required.

\subsection{Limitations of Cloud Evaluation}
  \label{sec:cloud/limitations}

As our cloud analysis re-uses the server-side measurement data from \autoref{sec:serverside/methodology}, it shares the same limitations discussed in \autoref{sec:serverside/limitations}.

While we evaluate deployment defaults (such as default-on and opt-out),
  capturing trends in cloud IPv6 adoption, 
  this analysis does not evaluate secondary differences.
For example, despite sharing the same default policy, Cloudflare has more IPv6 use than Akamai, both overall, and in direct comparisons.
This difference may reflect Akamai's longer operational history
  and pre-existing customer setups from before its default-on policy.
This suggests that additional factors---such as historical client configurations---may contribute to real-world cloud adoption differences.
We leave a more detailed investigation of these differences to future work.

\section{Conclusion}

This paper advocates for a more nuanced, non-binary view of IPv6:
  not just whether IPv6 is supported, but how much it is actually used in practice.
We apply this non-binary perspective through measurements at three levels of the ecosystem: users, services, and cloud providers, and show that binary metrics (such as ``IPv6 supported or not'') often miss key aspects of real-world IPv6 usage.

Our measurements show that even in dual-stack environments, IPv6 usage varies significantly with human activity, and many commonly accessed services remain IPv4-only.
At the service level, while over 40\% top websites now support IPv6 at the root, full IPv6 coverage across all embedded resources is rare.
Among cloud platforms, we find that tenant adoption is strongly influenced by the ease of enabling IPv6---platforms with IPv6-on-by-default policies see much higher uptake.

Based on our findings that highlight the limitations of binary adoption metrics, we call on the measurement community to assess IPv6 progress based on actual usage and resource-level coverage.
Moving toward full IPv6 adoption will require not just support, but seamless, default-on deployment from major applications, enterprises, third-party resources, and cloud providers.

\begin{acks}

This work is partially supported by the U.S.~National
  Science Foundation through 
  projects
  ``Privacy in Internet Measurements Applied To WAN and Telematics (PIMAWAT)''
   (CNS-2319409),
  ``A Traffic Map for the Internet''
   (CNS-2212480),
and BRIPOD (OAC-2530698).
We sincerely thank the volunteers who installed our measurement tool on their home routers, enabling us to collect the data used in this paper.
We also thank the anonymous reviewers and shepherd for their thoughtful feedback and suggestions.

\end{acks}

\printbibliography

\appendix
\section{Ethics}
    \label{sec:appendix/ethics}

Our user studies in this paper were IRB-reviewed 
  (USC IRB \#UP-24-00738)
and approved.
As part of IRB approval we documented our user interaction scripts,
  user population, and opt-out procedures.

For client-side data collection,
our measurement tool operates on household routers, recording the IP endpoints of traffic flows and the amount of data sent and received.
Since this information is generally considered private, we took several steps to protect user privacy:
The user's router 
  anonymized the user's identity and IP address
  before uploading data to our servers,
  scrambling the lower 8 bits of IPv4 addresses and the lower /64 of IPv6
with CryptoPAN~\cite{Xu02b}.
We we also transmit data from the user's router to our server securely with TLS,
  and store the data with access limited to the authors.

\section{MSTL Results}
  \label{sec:appendix/mstl-results}

\begin{figure*}
  \centering
  \includegraphics[width=0.8\linewidth]{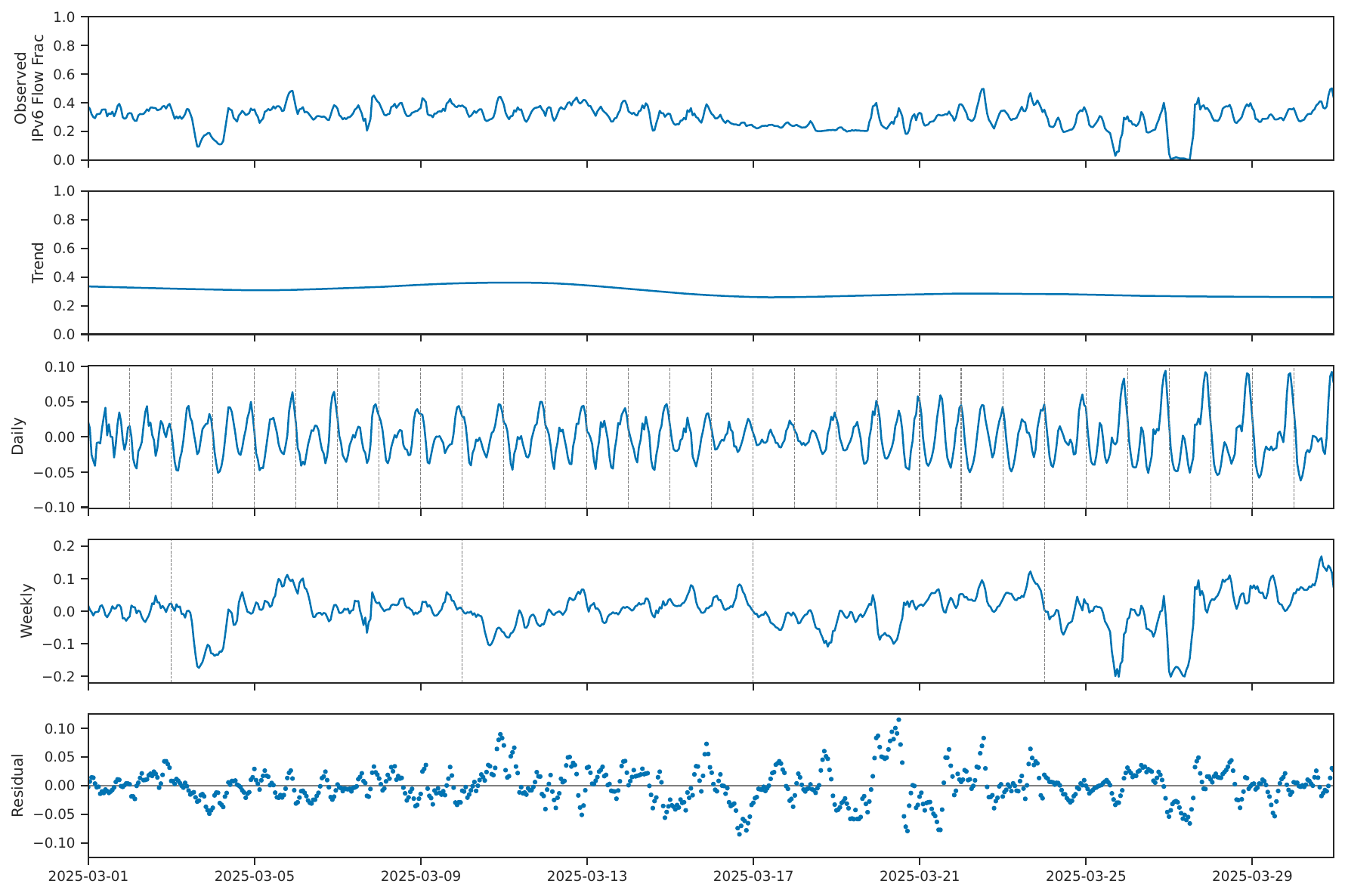}
  \caption{
      Fraction of IPv6 flows at residence A (top graph), with MSTL decomposition into trend, daily, weekly, and residual components.
  }
  \label{fig:appendix/mstl-residenceA-march-flows}
\end{figure*}

\begin{figure*}
  \centering
  \includegraphics[width=0.8\linewidth]{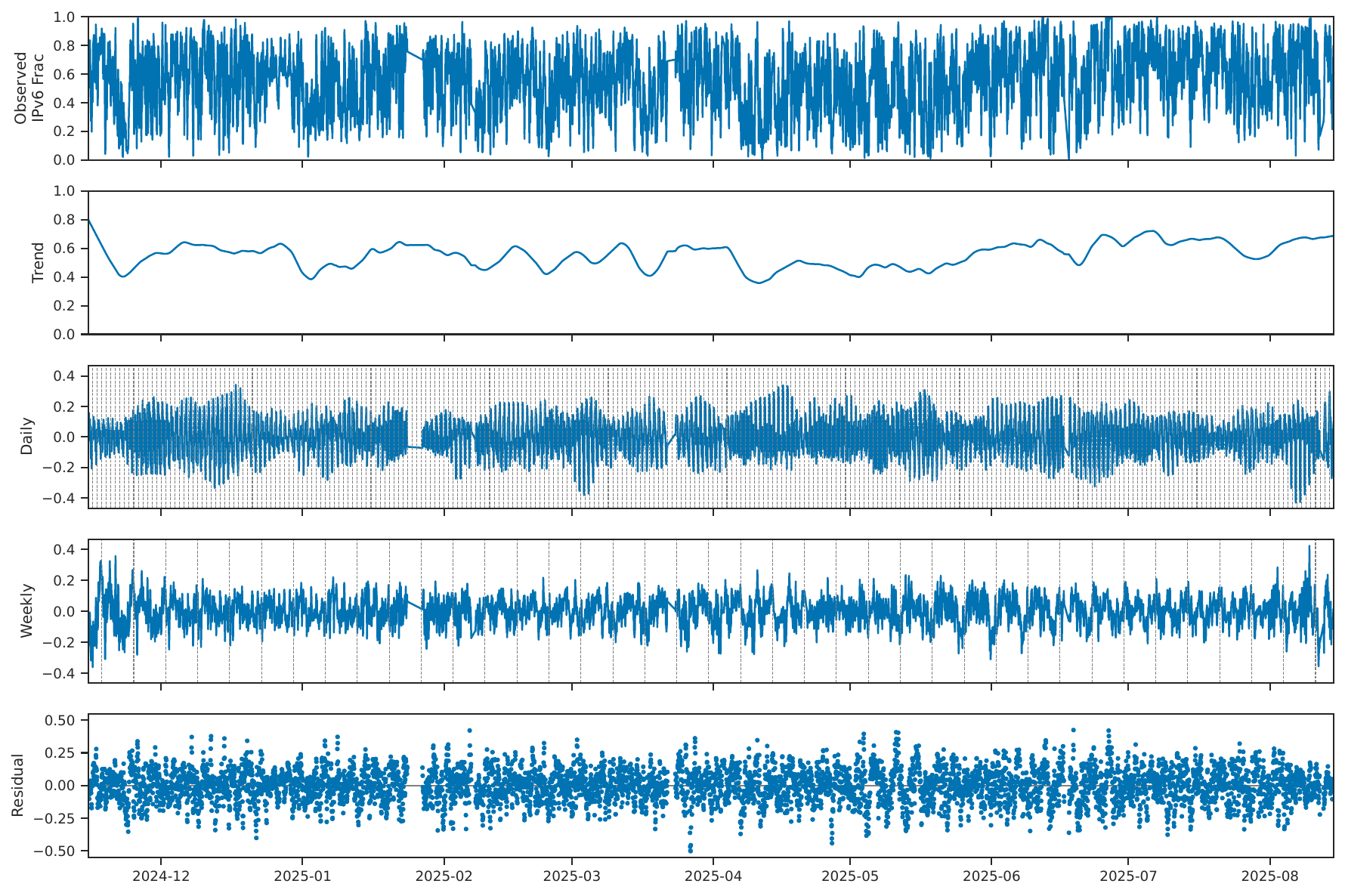}
  \caption{
      MSTL of daily IPv6 byte fractions at \ResRouter{B} over the full observation period.
  }
  \label{fig:appendix/mstl-residenceB}
\end{figure*}

\begin{figure*}
  \centering
  \includegraphics[width=0.8\linewidth]{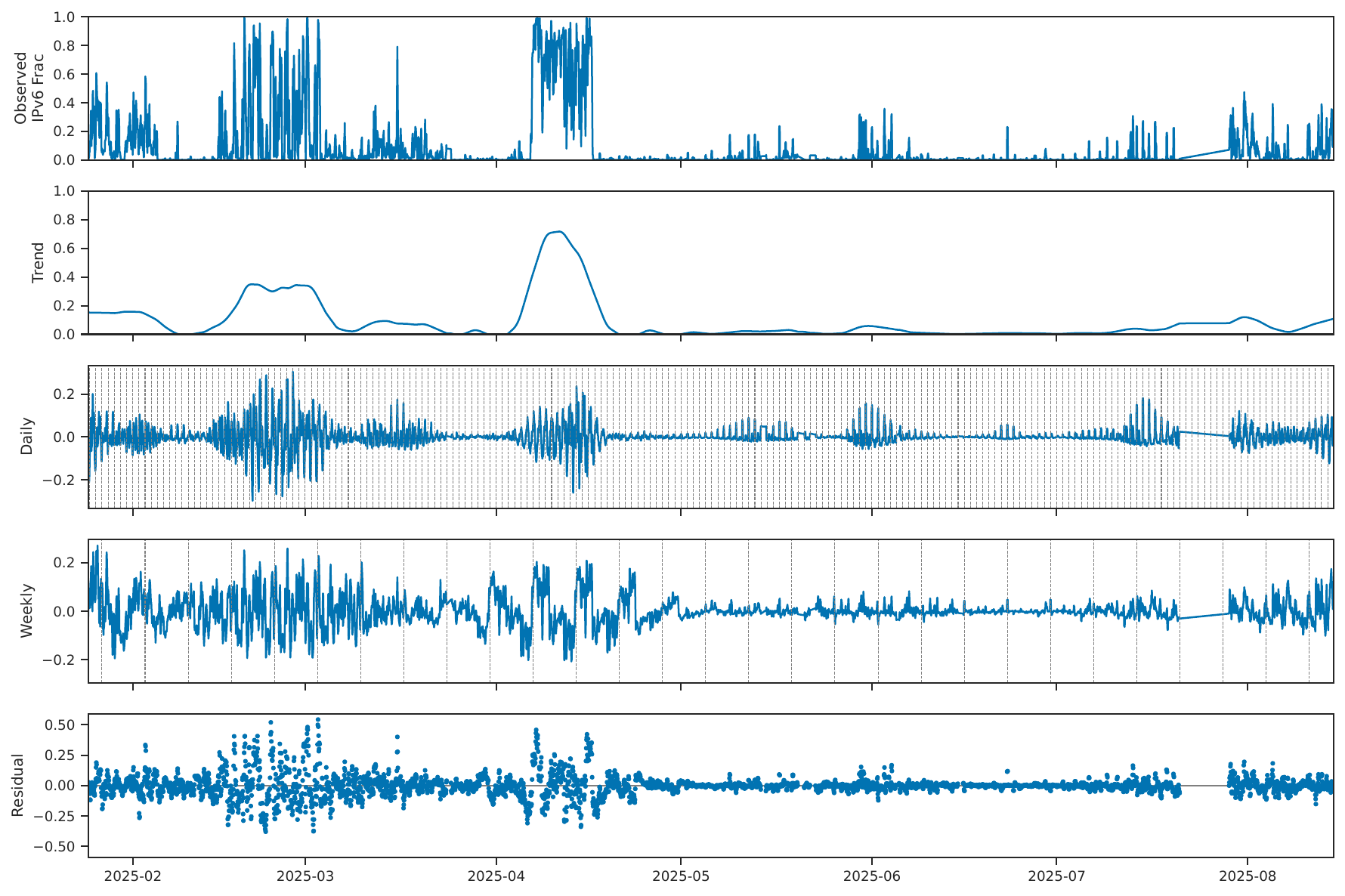}
  \caption{
      MSTL of daily IPv6 byte fractions over the six month observation period at \ResRouter{C}.
  }
  \label{fig:appendix/mstl-residenceC}
\end{figure*}

\autoref{fig:appendix/mstl-residenceA-march-flows} shows the MSTL decomposition of IPv6 flow fractions at \ResRouter{A} during March 2025.
\autoref{fig:appendix/mstl-residenceB} and \autoref{fig:appendix/mstl-residenceC} show the MSTL of daily IPv6 byte fractions at \ResRouter{B} and \ResRouter{C} over the full observation period.
Discussion of these results is in \autoref{sec:client/temporal}.

\section{Details about Other Residences}
  \label{sec:appendix/other_residences}

\begin{figure*}
  \centering

  \begin{subfigure}[t]{0.4\textwidth}
    \centering
    \includegraphics[width=\linewidth]{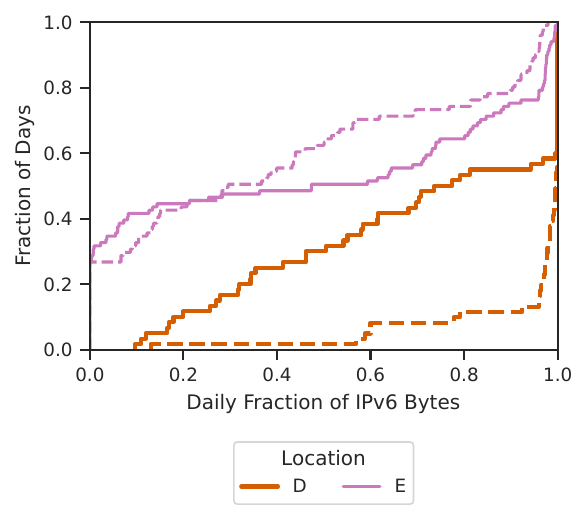}
    \caption{Bytes}
    \label{fig:appendix/cdf-v4v6split-DE-bytes}
  \end{subfigure}
  \hfill
  \begin{subfigure}[t]{0.4\textwidth}
    \centering
    \includegraphics[width=\linewidth]{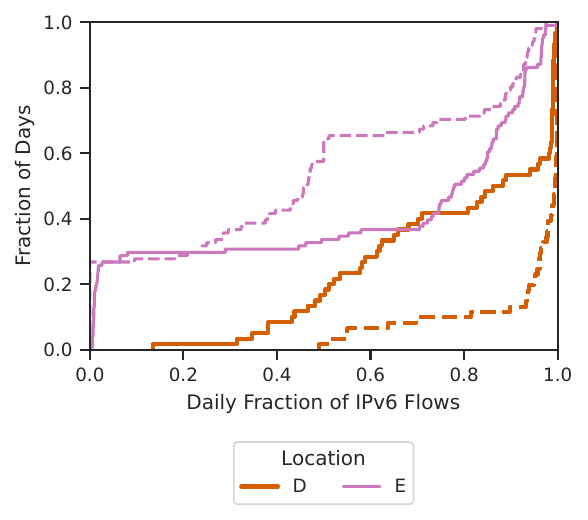}
    \caption{Flows}
    \label{fig:appendix/cdf-v4v6split-DE-flows}
  \end{subfigure}

  \caption{
    Fraction of per-day IPv6 bytes and flows (vs. IPv4) at Residences D and E, by external (solid lines) and internal (dashed lines).
  }

  \label{fig:appendix/cdf-v4v6split-DE}
\end{figure*}

To augment the results in \autoref{sec:client/v6fraction},  
  we provide the
  fraction of traffic and flows for our two lightest-traffic residences in
  \autoref{fig:appendix/cdf-v4v6split-DE}.
These residences show even greater varation over days than Residences A, B, and C.

\section{Details about Client-accessed Services}
   \label{sec:appendix/client_service_details}

\begin{figure*}
  \centering
  \includegraphics[width=\textwidth]{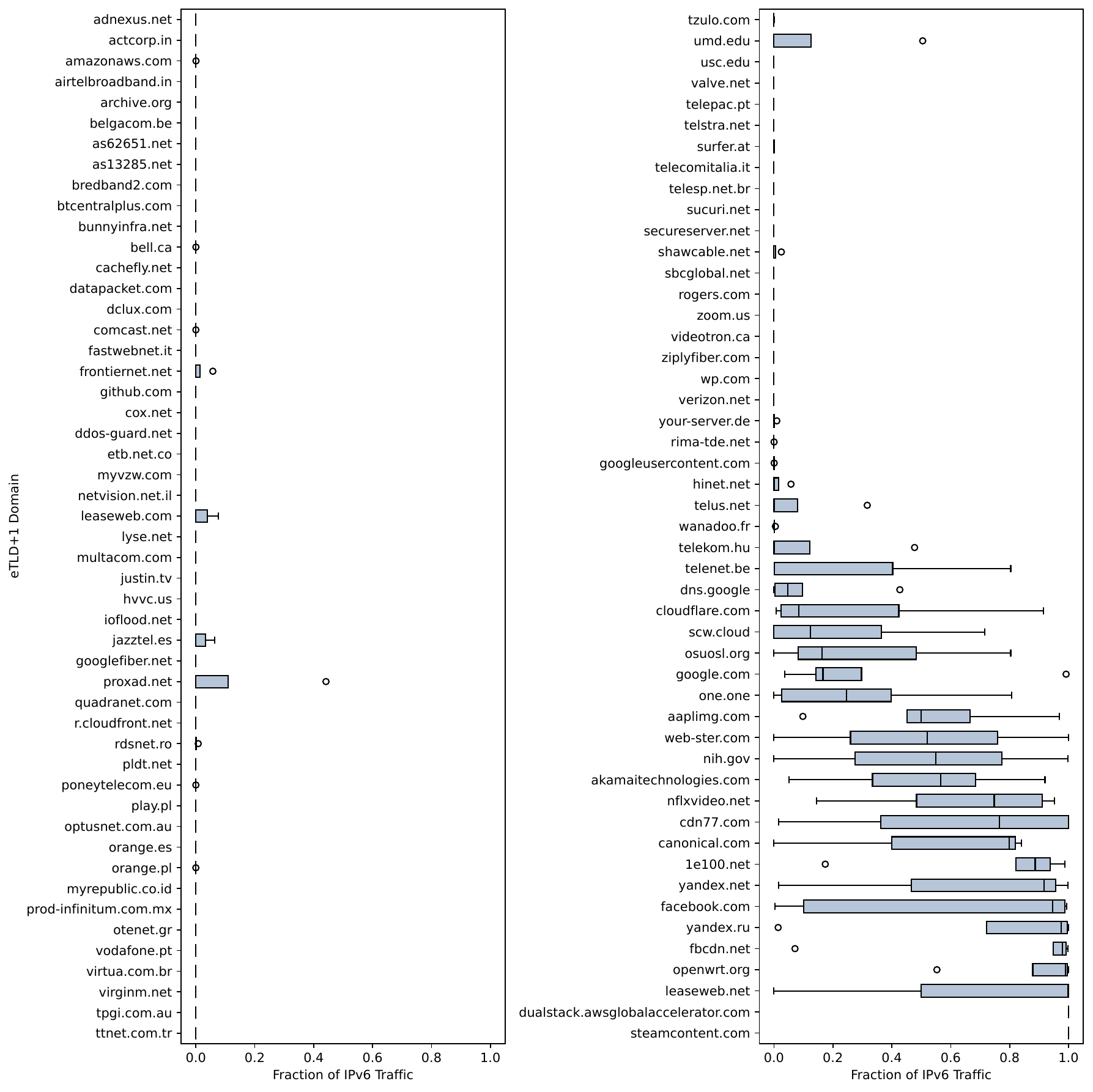}
  \caption{
    Distribution of IPv6 fractions for domains (eTLD+1) appearing at at least 3 locations and accounting for at least 100MB of traffic.
    Boxes show the interquartile range (25th-75th percentiles), and whiskers extend to 1.5$\times$IQR.
    Dots represent outliers.
  }
  \label{fig:appendix/boxplot-ipv6frac-domains}
\end{figure*}

\autoref{sec:client/leaders-laggards} reports how some services popular with clients
  lead or lag in IPv6 deployment.
We list specific domains we see in \autoref{fig:appendix/boxplot-ipv6frac-domains}.

\section{Server-side IPv4-only Resources}
    \label{sec:appendix/serverside-third-party-laggards}

\begin{figure*}
  \centering
  \includegraphics[width=0.9\linewidth]{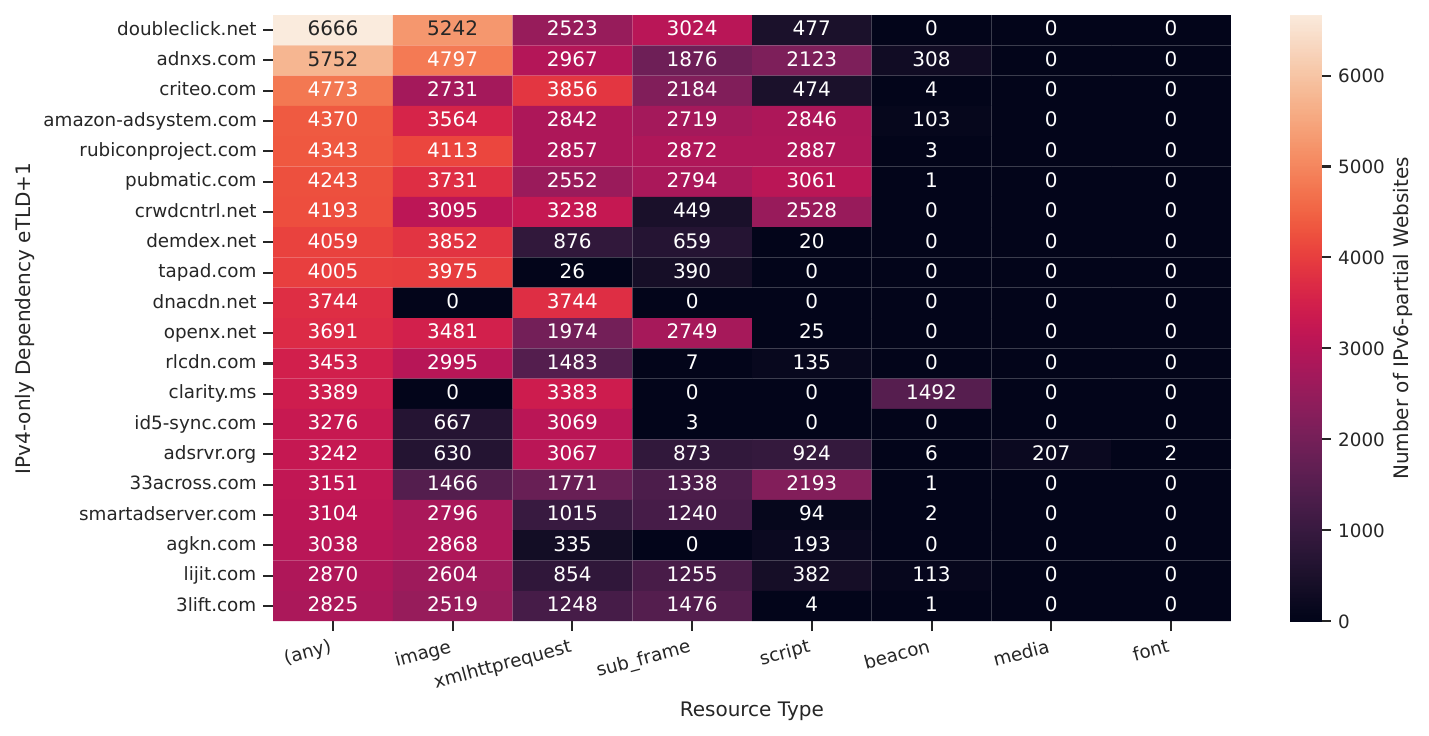}
  \caption{
    Heatmap showing the top 20 IPv4-only resources relied upon by IPv6-partial websites, organized by resource type.
  }
  \label{fig:appendix/serverside-third-party-laggards}
\end{figure*}

\autoref{sec:serverside/resource-domains}
  discusses the challenge of server-side IPv4-only resources,
  and \autoref{fig:appendix/serverside-third-party-laggards}
  lists specific IPv4-only resources we see.

\section{Specific Cloud Domains}
	\label{sec:appendix/cloud-domain-breakdown}

\autoref{sec:cloud/cloud-share} discusses which clouds show
  greater IPv6 use.
\autoref{tab:appendix/cloud-domain-breakdown}
  compares the top 15 clouds,
  how many domains we see hosted in each,
  and their IPv6 support.

\begin{table*}
  \centering
  \resizebox{0.7\textwidth}{!}{
    \begin{tabular}{lrrrrrrr}
      \toprule
      \multirow{2}{*}{\textbf{Cloud}} &
      \multirow{2}{*}{\textbf{\# of Domains}} &
      \multicolumn{2}{c}{\textbf{\# IPv4-only}} &
      \multicolumn{2}{c}{\textbf{\# IPv6-full}} &
      \multicolumn{2}{c}{\textbf{\# IPv6-only}} \\
      \cline{3-4} \cline{5-6} \cline{7-8}
      & & \textbf{Count} & \textbf{(\%)} & \textbf{Count} & \textbf{(\%)} & \textbf{Count} & \textbf{(\%)} \\
      \midrule
      \emph{Overall}            & 272\,964 & 153\,786 & 56.3\% & 114\,465 & 41.9\% & 4,713 & 1.7\% \\
      \midrule
      Cloudflare, Inc.          & 59\,106 & 8\,743  & 14.8\% & 50\,355 & 85.2\% & 8     & 0.0\% \\
      Amazon.com, Inc.          & 57\,856 & 42\,885 & 74.1\% & 14\,254 & 24.6\% & 717   & 1.2\% \\
      Google LLC                & 40\,735 & 13\,138 & 32.3\% & 27\,584 & 67.7\% & 13    & 0.0\% \\
      \midrule
      Akamai International B.V. & 10\,533 & 3\,654  & 34.7\% & 5\,305  & 50.4\% & 1\,574 & 14.9\% \\
      Fastly, Inc.              & 7\,739  & 5\,072  & 65.5\% & 2\,655  & 34.3\% & 12    & 0.2\% \\
      Microsoft Corporation     & 5\,480  & 3\,300  & 60.2\% & 2\,177  & 39.7\% & 3     & 0.1\% \\
      \midrule
      Akamai Technologies, Inc. & 5\,416  & 5\,211  & 96.2\% & 184    & 3.4\%  & 21    & 0.4\% \\
      Cloudflare London, LLC    & 3\,474  & 2\,899  & 83.4\% & 575    & 16.6\% & 0     & 0.0\% \\
      Hetzner Online GmbH       & 3\,303  & 2\,715  & 82.2\% & 576    & 17.4\% & 12    & 0.4\% \\
      \midrule
      OVH SAS                   & 3\,134  & 2\,714  & 86.6\% & 408    & 13.0\% & 12    & 0.4\% \\
      Hangzhou Alibaba [...]    & 3\,003  & 2\,388  & 79.5\% & 608    & 20.2\% & 7     & 0.2\% \\
      Datacamp Limited          & 2\,885  & 1\,742  & 60.4\% & 1\,142  & 39.6\% & 1     & 0.0\% \\
      \midrule
      DigitalOcean, LLC         & 1\,899  & 1\,718  & 90.5\% & 175    & 9.2\%  & 6     & 0.3\% \\
      Incapsula Inc             & 1\,363  & 1\,313  & 96.3\% & 48     & 3.5\%  & 2     & 0.1\% \\
      BUNNYWAY, [...]           & 1\,316  & 6      & 0.5\%  & 0      & 0.0\%  & 1\,310 & 99.5\% \\
      \bottomrule
    \end{tabular}
  }

  \caption{Count and percentage breakdown of domains by cloud provider.}
  \label{tab:appendix/cloud-domain-breakdown}
\end{table*}

\end{document}